\documentclass[graybox,natbib,nosecnum]{svmult}
\bibpunct{(}{)}{;}{a}{}{,} 

\pdfoutput=1   

\usepackage{mathptmx}       
\usepackage{helvet}         
\usepackage{courier}        
\usepackage{type1cm}        

\usepackage{makeidx}         
\usepackage{graphicx}        
\usepackage{multicol}        
\usepackage[bottom]{footmisc}
\usepackage[normalem]{ulem}	
\usepackage{hyperref}  

\usepackage{soul}   

\newcommand{\hbindex}[1]{\hl{#1}\index{#1}}  

\makeindex             

\begin{document}

\title*{Special cases : moons, rings, comets, trojans}
\author{Juan Cabrera, Mar{\'i}a Fern{\'a}ndez Jim{\'e}nez, Antonio Garc{\'i}a Mu{\~n}oz, Jean Schneider}
\institute{Juan Cabrera \at Institut f\"ur Planetenforschung, Deutsches Zentrum f\"ur Luft- und Raumfahrt, Rutherfordstr. 2, Berlin, Germany, \email{juan.cabrera@dlr.de}
  \and Mar{\'i}a Fern{\'a}ndez Jim{\'e}nez \at Institut f\"ur Planetenforschung, Deutsches Zentrum f\"ur Luft- und Raumfahrt, Rutherfordstr. 2, Berlin, Germany, \email{maria.fernandez.jimene@gmail.com}
  \and Antonio Garc{\'i}a Mu{\~n}oz \at Zentrum f\"r Astronomie und Astrophysik, TU Berlin, Hardenbergstraße 36, D-10623 Berlin, Germany, \email{garciamunoz@astro.physik.tu-berlin.de}
  \and Jean Schneider \at LUTH, Observatoire de Paris, CNRS, Université Paris Diderot; 5 place Jules Janssen, 92195 Meudon, France, \email{jean.schneider@obspm.fr}
}
%
%
\maketitle

\abstract{Non-planetary bodies provide valuable insight into our
  current understanding of planetary formation and evolution.
  Although these objects are challenging to detect and characterize,
  the potential information to be drawn from them has motivated
  various searches through a number of techniques.
  Here, we briefly review the current status in the search of moons,
  rings, comets, and trojans in exoplanet systems and suggest what
  future discoveries may occur in the near future.}

\section{Introduction}

It is not original admitting that it is complicated to make accurate
predictions about the future. 
There is a quote, attributed to Napoleon Bonaparte, warning against
insisting too much in having the full control of the present status of
a problem and a detailed plan for the future developments before
starting to do the work
(\emph{Celui qui, au d{\'e}part, insiste pour savoir o{\`u} il va,
  quand il part et par o{\`u} il passe n'ira pas loin).}
However, considering the importance of the investment required to
answer certain scientific questions, it is mandatory to have a
realistic idea of the likelihood of success of the research. 
These considerations apply to the topic of this chapter: the search
and characterization of moons, rings, comets, and trojans in
exoplanetary systems.

We know from the solar system that satellites and ring systems, minor
planets, and comets provide meaningful insights into the processes of
planetary formation and evolution~\citep{depater2015}. 
Many of them are interesting objects of research on their own, in
particular considering their prospects of habitability. 
However, in view of the difficulty of categorically proving, or ruling
out, the existence of life on other worlds in our solar system
\citep[i.e.][]{waite2017}, one can rightfully wonder about the
possibilities of actually finding life in extrasolar systems
(for a summary of the technical difficulties, see
\citealt{schneider2010}; for a recent review on habitability, see
\citealt{cockell2016}).

In the following Sections we briefly discuss what to expect from near
future researches about extrasolar systems of moons, rings, comets,
and trojan minor planets. 

\section{Exomoons}

In this volume there is an excellent review on exomoons and ring
detections by R. Heller, therefore we have orientated this chapter
towards complementary aspects.
Additionally, we refer the reader to some recent reviews
on the topic
by~\citet{barr2017a,kipping2014b,schneider2015,sinukoff2013}.
In order not to dwell long on topics already addressed by previous
reviews, we will only briefly discuss processes of exomoon formation
and evolution before addressing the present status of discoveries and
our expectations for the future.

The research on the processes leading to the formation of exomoons
has benefited from studies applied to the solar system
\citep[see][]{heller2015a,miguel2016,ogihara2012,crida2012}.
However, exomoons are expected to be found in different environments
depending on the details of their evolution in the
disk~\citep{fujii2017}, the outcome of scattering
processes~\citep{gong2013}, capture~\citep{ochiai2014} or  
collisions~\citep{barr2017b}, to name a few.
Exotic situations where planets are ejected from the planetary system
conserving their moons have been mentioned, though their detection is
extremely challenging in this configuration~\citep{laughlin2000}. 

Moons exist between the Roche lobe and the Hill radius of their
host planets~\citep{murray2000}. 
There are numerous studies researching the dynamical stability and the
tidal evolution of moons
\citep{adams2016b,barnes2002,debes2007,domingos2006,donnison2010,hong2015,namouni2010,payne2013,sasaki2012,sasaki2014}.
Consequently, their final configuration will depend on planetary
processes like migration~\citep{spalding2016},
photoevaporation~\citep{yang2016}, or tidal
interactions~\citep{cassidy2009}, all of them open to a certain
degree of interpretation today.
Rather than a disadvantage this is an encouragement to study moons,
as they could provide useful measurable constraints.
However, the expected diversity requires observational support to
understand the relative impact of the different processes proposed. 

Habitability is another important reason to look for
exomoons~\citep{kaltenegger2010a,lammer2014}.
There are interesting processes that are exclusive of these systems
and that deserve specific attention.
They include tidal
interactions~\citep{dobos2017,forgan2013,heller2012,scharf2006},
planetary illumination~\citep{forgan2014}, amount of volatiles
depending on the formation and migration
mechanisms~\citep{heller2015a,heller2015b}.
Finally, the possible presence of moons might impact the
interpretation of biosignatures~\citep{rein2014,li2016}.

\subsection{Current status of detections}

There is no known reason preventing exomoons from existing and we
expect them to be present in many different configurations. 
Therefore, all detection methods applied to
exoplanets~\citep{wright2012} have also been extended to detect
exomoons with a varying degree of predicted success rate. 
The different detection methods are excellently described in Heller's
review in this volume and we will refer the reader to that text for an 
overview.

Almost twenty years ago there were high expectations on the detection 
possibilities of exomoons with space-borne facilities like
Hubble~\citep{brown2001}, \emph{CoRoT}~\citep{sartoretti1999}, or
\emph{Kepler}~\citep{szabo2006}; but also with microlensing~\citep{gaudi2003}
or direct imaging~\citep{cabrera2007}.

However, there is no uncontroversial detection of an exomoon as we
write this lines in August 2017.
The situation might change soon, as we will see in the next section. 
Which have been the difficulties encountered?

Photometric detections during transit and occultation are
challenging, even with the latest
instrumentation~\citep[e.g. see][]{dobos2016},
and have two main practical limitations: stellar activity and
instrumental systematics.
A paradigmatic example is the transit of TrEs-1b~\citep{rabus2009b},
whose Hubble light curve can be interpreted as a two planet system or
as the passage of the transiting planet over an active region on the
stellar surface.
The same difficulty has been encountered by other systematic
studies~\citep{lewis2015}.
More challenging is the occurrence of instrumental systematics
mimicking the effects of moons, like the case of 
Kepler-90g~\citep{kipping2015a}.
Instrumental systematics can be very difficult to eliminate, even in 
presence of large amounts of data with high photometric
quality~\citep[see, for example,][]{gaidos2016}.

Therefore, the problem is not necessarily the detection but rather the
unique interpretation of the measurement as being caused by an
exomoon.
There is indeed a number of processes that can lead to comparable
observational effects but do not involve exomoons.
In this respect, the transit timing variation (TTV) method is
considered promising, as it can solve part of the degeneracies
intrinsic to photometric
detections~\citep[see][and references therein]{lewis2014}. 
However, the transit timing variations (TTVs) of Kepler-46b
(KOI-872b) can be interpreted as an additional planet in the
system or as a moon~\citep{nesvorny2012a}.
TTVs are also strongly affected by stellar
activity~\citep{barros2013,lewis2013} and
systematics~\citep{szabo2013}.
As a result of these limitations, there are presently many systems
studied~\citep{weidner2010,kipping2013a,kipping2013b,kipping2014b,kipping2015b,hippke2015a,kane2017},
but no claimed detection of exomoon.

Microlensing surveys have suffered from similar difficulties in the 
interpretation of the observations with the added challenge of the
reproducibility of the measurements~\citep{bennet2014,skowron2014}.

\begin{figure}
\includegraphics[%
    width=\linewidth,%
    height=0.5\textheight,%
    keepaspectratio]{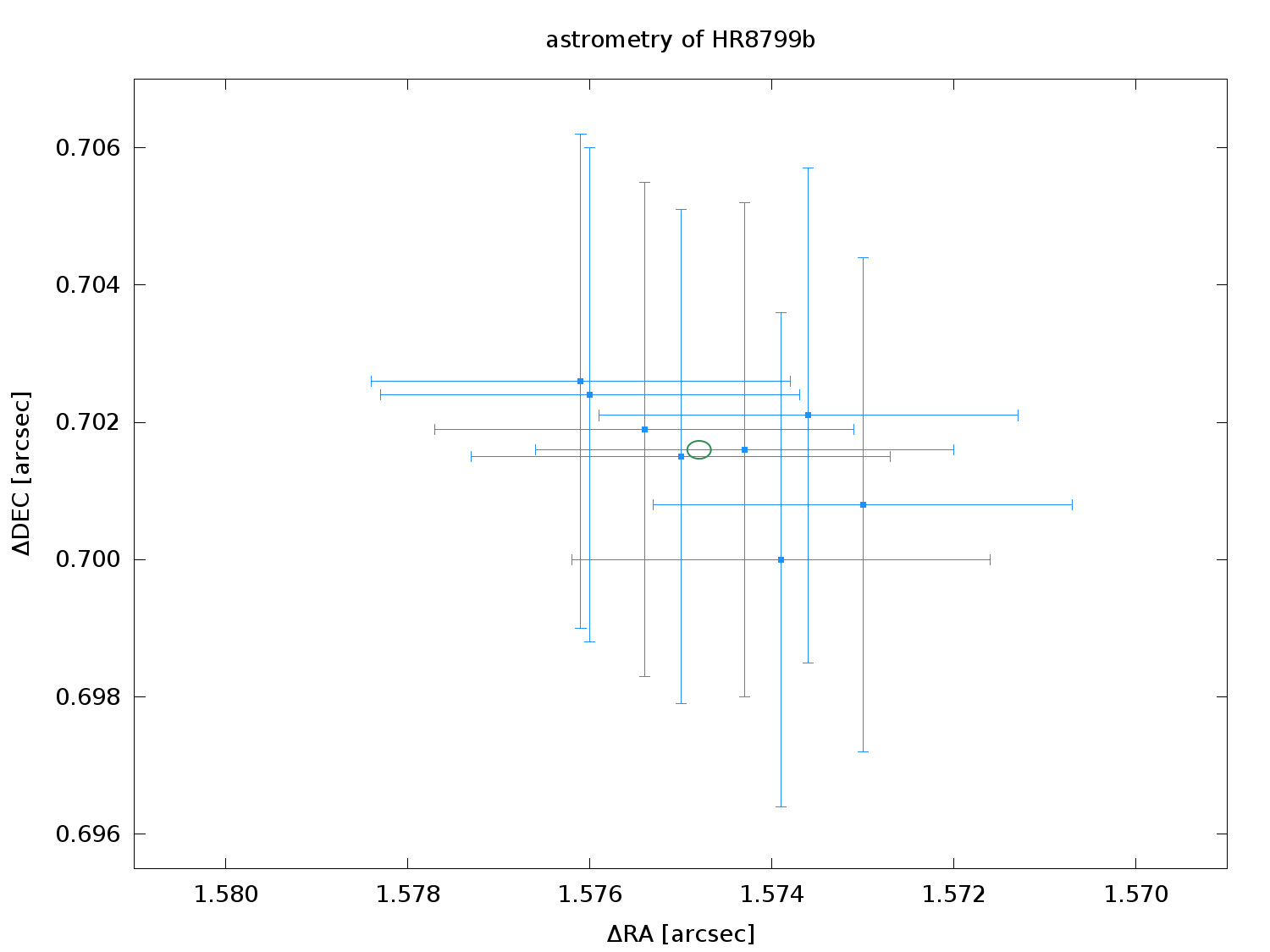}
\caption{Comparison between the astrometric measurements obtained for
  HR8799b by~\citet{wertz2017} and the expected photocenter
  motion for a Jupiter-Saturn binary planet observed at 39
  pc. See~\citet{cabrera2007} for details on how has been calculated
  the expected motion of the photocenter.}
\label{fig:photocenter}
\end{figure}

Ten years ago, we estimated the possibility of detecting moons around
direct imaged planets measuring the reflex motion of the moon around
the planet considering photon noise~\citep{cabrera2007}.
The required precision of the astrometric measurement of the position
of the planet was in the range from microarcsec to few milliarcsec.
Unfortunately, it is more likely that these observations are actually
limited by speckle noise and systematic uncertainties in the
astrometric position of the host star rather than photon noise.
However, precisions of milliarcseconds are currently within
reach~\citep{wertz2017}, though no moon has been claimed
yet~(see Fig.~\ref{fig:photocenter}). 

\subsection{Expectations for the future}

\begin{figure}
\includegraphics[%
    width=\linewidth,%
    height=0.5\textheight,%
    keepaspectratio]{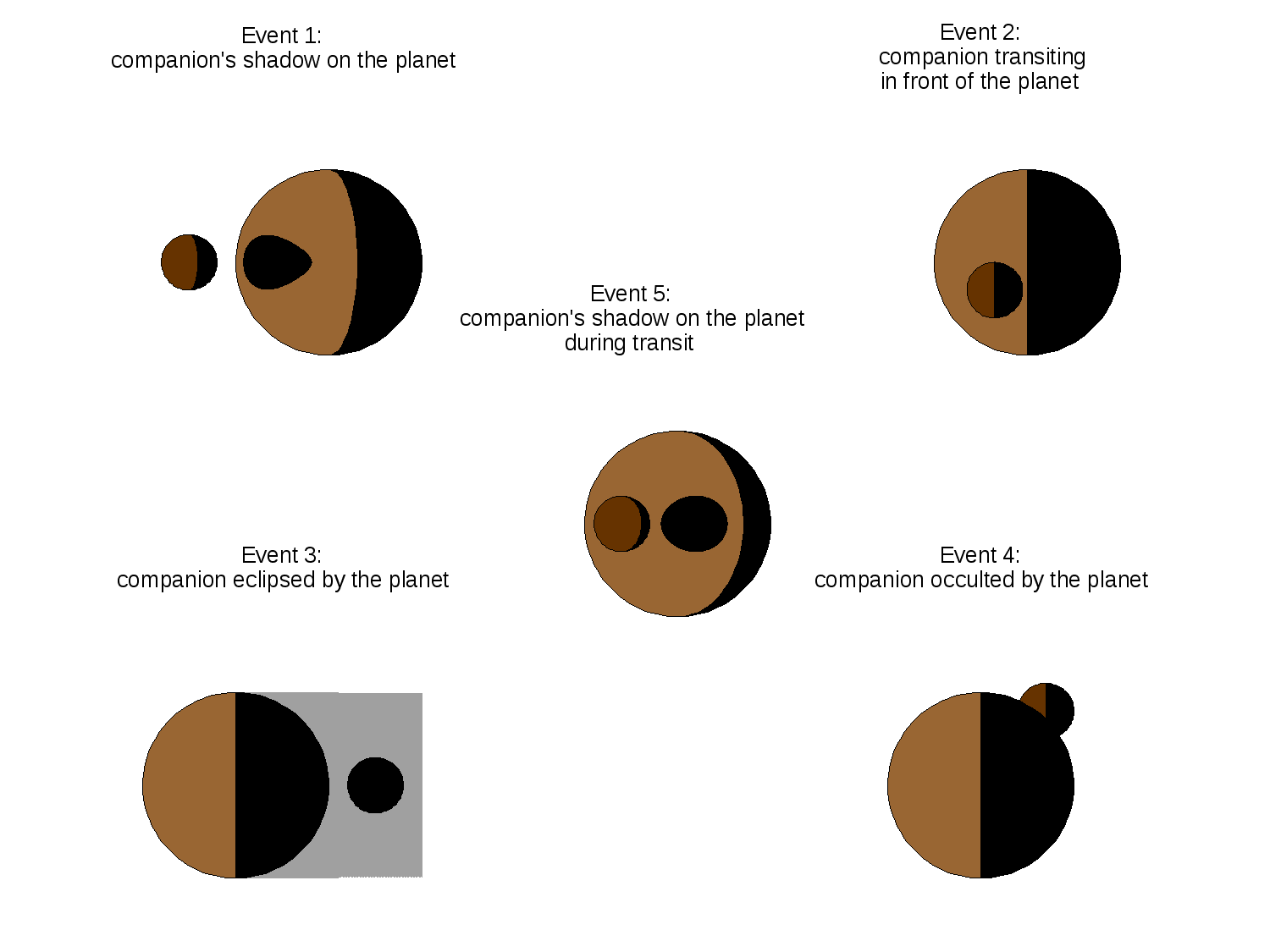}
\caption{Planet/moon events. As a satellite orbits its host planet,
  there are different events that can be observed in the system,
  depending on the relative configuration of the planet, satellite,
  and observer.}
\label{fig:events}
\end{figure}

The photometric detection of moons is theoretically possible with
space-borne photometry delivered by missions such as \emph{CoRoT} and
\emph{Kepler} or with microlensing surveys, but so far no detection
has been secured. 
Soon the next generation of exoplanet space-borne facilities,
including \emph{CHEOPS}~\citep{simon2015},
\emph{TESS}~\citep{ricker2015}, and \emph{PLATO}~\citep{rauer2014},
will expand on the \emph{CoRoT} and \emph{Kepler} legacy.
The difficulty in the detection of exomoons will be the same, but
these missions will observe brighter stars, easier to characterize,
and will benefit from the experience of the previous surveys.
For example, one wonders about the follow-up of the TTVs of
\emph{Kepler} candidates with \emph{PLATO}, collecting a baseline of
observations longer than ten years.
\emph{PLATO} has a smaller collecting area than \emph{Kepler}, but a
higher cadence, so the TTV accuracy will be close enough to
make meaningful comparisons.
These new missions might be able to definitely settle the nature
of some of the candidates proposed today.

An important limitation of transit photometry arises from the fact
that a system is in transit only a very small fraction of its orbit
(0.15\% of the time for the Earth around the Sun).
The relative scarcity of the data and the difficulty of reproducing
the observations, as the moon changes it relative phase from transit
to transit, will not improve for the new missions.
A possibility that might still have a chance are binary planets, which
have a very distinct transit signature of larger amplitude.
Though we know they are not common, they are known to exist in certain
configurations~\citep{nielsen2013,best2017,han2017}.

Despite all the mentioned difficulties, in July
2017~\citet{teachey2018} announced the presence of a possible exomoon
around the planet Kepler-1625b.
If confirmed, it would be a sensational discovery culminating the
efforts of the HEK (Hunt for Exomoons with Kepler)
team~\citep{kipping2012c}.
However, the authors remain cautious about the nature of the candidate
and warn the community about the limited amount of existing
observational evidence on this target.
There are additional observations scheduled emd pf October 2017 with
the potential to confirm the presence of the candidate.
Unfortunately, this text will have to go into publication before the
results of these observations are known, but we keep our fingers
crossed.

In the near future, another breakthrough might come though from
directly imaged planets~\citep[see the review by][]{bowler2016}.
As mentioned earlier, current facilities can reach the
milliarcsecond precision required to start sampling the existence of
massive moons (see Fig.~\ref{fig:photocenter}).
There are important synergyes with missions like
\emph{Gaia}~\citep{brown2016GaiaDR1}, which will probe the parameter
space for giant planets beyond the snow-line, potentially observable
with current high-contrast adaptive optics facilities.
\emph{Gaia} planet yield is expected to outnumber
the current sample known~\citep{casertano2008}.
The reflex motion of planets, which allows to measure the moon's mass,
and the next step, studying planet-moon occultation
events~\citep[see][]{schneider2015}, are promising methods to
characterize exomoons.
The reflex motion is observable during the whole orbit and the
planet-moon events occur up to 5 times per moon orbit (see 
Fig.~\ref{fig:events}), which is in the order of up to a few tenths of
days, giving considerable advantage in comparison to photometric
transits. 
Observing the 5 types of events shown in Fig.~\ref{fig:events} is only
possible if the relative orientation of orbit of the satellite is
favorable to the observer. 
However, two of the events, when the satellite casts its shadow on
the planet and when the satellite is occulted by the shadow of the
planet, only depend on the relative orientation of the satellite orbit
with the orbital plane of the planet.
If the orbital plane of the satellites has a low inclination with
respect to the ecliptic, these events will be visible, regardless of
the orientation towards the observer.

Though elusive, exomoons are fundamental in the study of the
processes of planetary formation and evolution and, furthermore,
provide a rich ground for studies constraining the processes of
planetary formation and evolution and provide a great 
opportunity to study habitable systems.
Given the new facilities that will become available in the coming
years and the expertise accumulated, there are good reasons to remain
optimistic and hope that the next 10 years will be more fruitful than
the last two decades.

\section{Rings}

All the giant planets in the solar system are surronded by systems of
rings, though they have very different properties.
The processes that affect the stability and evolution of rings
involve tidal forces, dynamical interactions with moons, resonances,
spiral waves, radiation pressure, and interactions with charged
particles~\citep{depater2015}, making them a very rich field for
research. 

We bring up in this section rings and disks around planets, but there
are ring-structures everywhere in the Universe within an unimaginable
range of sizes~\citep[see, for example, the review by][]{latter2017}.

For the detection methods of rings around extrasolar planets, we will
refer to the review in this very same volume by R. Heller.
Regarding formation mechanisms, see~\citet{zanazzi2017} and
references therein.

\subsection{Current status of detections}

In contrast to exomoons, there are several detections claimed in the
literature, though not all of them completely undisputed.  

One example is the indirect detection of a ring system around the  brown
dwarf G~196-3~B~\citep{zakhozhay2017}.
A ring system with properties resembling those around Jupiter or
Neptune, but with a very different age and on different environment,
can satisfactory reproduce the colors of this target.

A hypothetical ring system that can make its \emph{premi{\`e}re} in
2017 would be the one around $\beta$~Pictoris~b~\citep{lagrange2010},
a giant planet orbiting a 10 million year old star with an orbital
period of about 35 years.
The large semi-major axis makes the transit probability meager, but it
might actually be transiting~\citep{lecavelier2016,wang2016} and a
campaign has been orchestrated to characterize the Hill sphere of the
planet as it crosses the stellar disk from
Earth~\citep{kenworthy2017}.

And there is the unusual, from our solar system perspective, system of
rings proposed around
J1407~\citep[1SWASP~J140747.93-394542.6][]{mamajek2012,rieder2016},
which is described in Heller's chapter in this volume.
We will mention it again in the next section.

Despite the wealth of giant planets at large orbital periods,
including several Jupiter-analogs, found by different
surveys~\citep{bedell2015,diaz2016b,esposito2013,foremanmackey2016,kipping2014c,kipping2016b,uehara2016},
space missions like \emph{CoRoT} and \emph{Kepler} have not yielded
any report of exorings so
far~\citep[see][]{heising2015,lecavelier2017,turner2016}.

The paper by~\citet{aizawa2017} deserves special attention as it
carefully shows that the limitations of the photometric method with 
current surveys are not dictated by the photon noise, but by the
residuals of systematic noise sources and the interpretation of the
results, as it was the case for exomoons previously described.

Overcoming the limitations of photometric searches, there are
alternative techniques like high resolution
spectroscopy~\citep{santos2015} and direct imaging.

\subsection{Characterization of rings with direct imaging}

WFIRST-AFTA~\citep{spergel2015wfirst} is an observatory of NASA
devoted to the study of dark matter, infrared astrophysics,
and extrasolar planets.
It is currently in its Phase A, undergoing the study of mission
requirements. 
The mission concept is based on a 2.4 m telescope with a large
field-of-view and is equipped with a wide field instrument and a
coronograph. 
The on-board spectrograph will foreseeably provide a contrast of
$10^{-9}$ and an inner working angle of $3\lambda/D$ at 430 nm.
Such a performance will enable the characterization of the atmospheres
of directly imaged exoplanets~\citep{greco2015}.

We have used a numerical model based on previous work
by~\citet{arnold2004} to simulate the integrated light curve of ringed
planets that could be observed with WFIRST. 
The exercise intends to elucidate the effects that exorings would
have in the phase curve and spectra of an exoplanet observed via
direct imaging, thereby drawing conclusions on the planet atmosphere
and the planet size.
The planets are assumed at orbital distances of 1-10 AU from their
star and 10 pc away from the solar system.

The numerical model considers mutual shadow of the planet on the ring
and the ring on the planet.
The shadow of the planet on the ring and the occultation are also
taken into account.
However, mutual reflection and shadow of the ring on the planet are
neglected.
The code accepts elliptical orbits and rings with fixed inner and
outer radius.
The planet is assumed to scatter starlight as a Lambertian
sphere~\citep{lester1979}, and the rings are assumed to be planar. 
At the ring, only single scattering is considered.

There are nine parameters that define the planet-ring
system geometry of the model: the planetary radius $R_p$, the
inclination of the orbital plane $i$, the wavelength dependent
planetary albedo $A_p$, the ring’s optical thickness $\tau_R$, the
ring's inner and outer radius $R_\mathrm{in}$ and $R_\mathrm{out}$, the single 
scattering albedo of the ring $\omega_o$, the ring's plane
inclination $i_R$, and the ring's plane intersection with
the orbital plane $\lambda_R$~\citep{arnold2004}.

\begin{figure}
\includegraphics[%
    width=\linewidth,%
    height=0.5\textheight,%
    keepaspectratio]{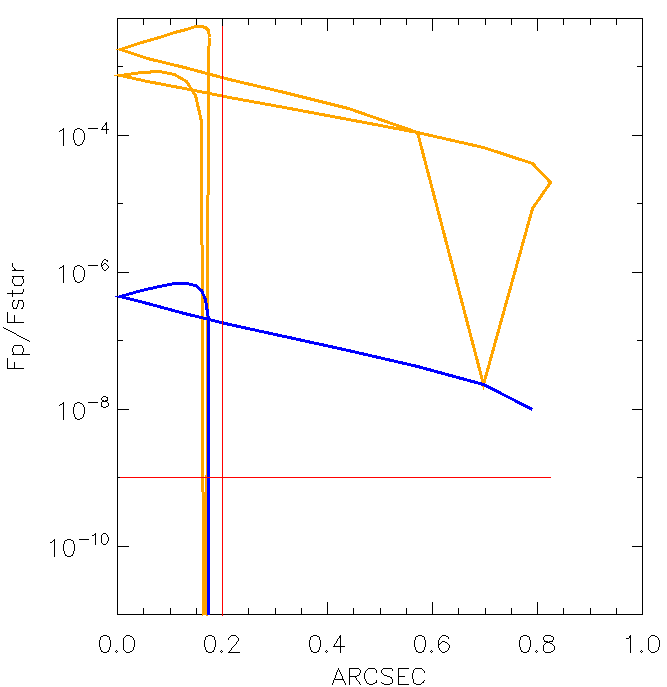}
\caption{Contrast vs. inner working angle for a J1407-like system
  compared to the WFIRST detection limits for a planet without rings
  (dark blue) and with rings (orange). The detectability improves
  moving from left to right and from bottom to top.
  The red horizontal line represents the planned $10^{-9}$ contrast
  limit of WFIRST. Correspondingly, the red vertical line represents a
  tentative limit of the inner working angle possibilities.} 
\label{fig:j1407iwa}
\end{figure}

\begin{figure*}
  \centering
  \begin{minipage}[t]{0.48\textwidth}
    \includegraphics[%
    width=\linewidth,%
    height=0.5\textheight,%
    keepaspectratio]{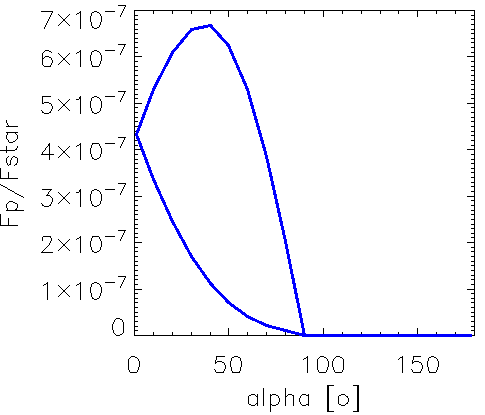}
  \end{minipage}
  \begin{minipage}[t]{0.48\textwidth}
    \includegraphics[%
    width=\linewidth,%
    height=0.5\textheight,%
    keepaspectratio]{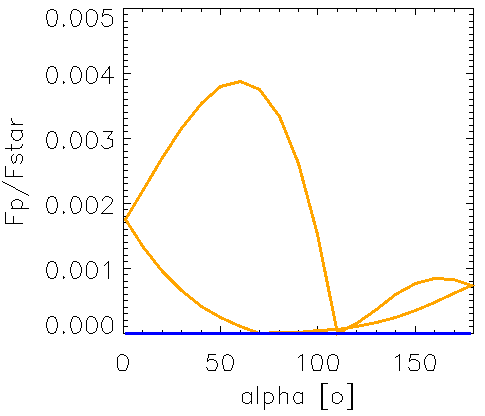}
  \end{minipage}
    \caption{Phase curves for a J1407-like system. Left: for a system
      without rings. Right: for a system with rings, note the change
      in the vertical scale. In the latter, the shape of the phase
      curve is dominated by the size of the rings and their relative
      orientation to the observer.}
    \label{fig:j1407refl}
\end{figure*}

As a reference for the capabilities of WFIRST we have made simulations
of a ring system like J1407.
To facilitate the study of this system with the planned specifications
of WFIRST's inner working angle, we assumed that the system is located
at 10 pc from Earth, rather than the actual 128 pc. 
The values of the parameters used in the simulation are shown in
Table~\ref{table:j1407ring} and the results are shown in
Figs.~\ref{fig:j1407iwa} and~\ref{fig:j1407refl}.

A system of rings as that proposed for J1407 reflects a
significant amount of the stellar light and produces a signal several 
orders of magnitude larger than the planet itself.
This unique case therefore opens the possibility of spectroscopically
investigating exorings without the interfering effect of the planet.  
Less extreme situations such as enabled by Saturn-like exoplanets will
show signals that blend the ring and planet contributions.
This blending will dilute the main absorption features in the planet
atmosphere, thereby complicating its analysis. 

\begin{table}
\caption{Parameters used in the simulation of the ring system.}
\label{table:j1407ring}
\begin{tabular}{p{2cm}p{2.5cm}p{3cm}}
\hline\noalign{\smallskip}
Element & Parameter & Value  \\
\noalign{\smallskip}\svhline\noalign{\smallskip}
Star    & mass            & 0.9M$_\mathrm{Sun}$       \\
        & distance        & 10 pc                   \\
\noalign{\smallskip}\svhline\noalign{\smallskip}
Planet  & $R_p$           & 1.46 $R_\mathrm{Jupiter}$ \\
        & semi-major axis & 5 au                    \\
        & eccentricity    & 0.65                    \\
        & $i$             & 89$^\circ$               \\
        & $A_p$           & Jupiter                 \\
\noalign{\smallskip}\svhline\noalign{\smallskip}
Ring    & $\tau_R$        & 0.5                     \\
        & $R_\mathrm{in}$  & 0.25 $R_\mathrm{Hill}$    \\
        & $R_\mathrm{out}$ & $R_\mathrm{Hill}$         \\
        & $i_R$           & 13$^\circ$               \\
        & $\lambda_R$     & 70$^\circ$               \\
\noalign{\smallskip}\hline\noalign{\smallskip}
\end{tabular}
\\The albedo values of Jupiter as a function of the wavelength are taken
from~\citet{karkoschka1994}. 
\end{table}

\section{Comets}

Comets are some of the largest structures in the solar system, if one
accounts for the extension of their
tails~\citep[i.e.][]{neugebauer2007}.
However, the very low density of the extended tails makes their
detection and characterization challenging outside the solar system.
Nevertheless, this difficulty didn't stop observers from tying to
observe the extended, low density, exospheres of extrasolar planets
soon after their
discovery~\citep{rauer2000,vidalmadjar2003,lecavelier2010,haswell2012,ehrenreich2012,ehrenreich2015,poppenhaeger2013}.
The evaporation of giant planets has been followed by the detection of
disintegrating small planets which display tails similar to
comets~\citep{rappaport2012,rappaport2014,sanchisojeda2015,vanderburg2015},
which in some cases has allowed the characterization of the properties
of the particles in the
tail~\citep{vanlieshout2014,vanlieshout2016,zhou2016,alonso2016,rappaport2016}. 
Of interest are also the transient signatures recently discovered
around the young star RIK-210, though their interpretation is not so
straightforward~\citep{david2017}. 

The discovery of comets with photometric transit surveys has been more
difficult that what was originally expected~\citep{lecavelier1999} for
possibly the same reasons that have been already described above.
There have been clear detections in the circumstellar disk of
young stars~\citep{kiefer2014a,kiefer2014b,eiroa2016,marino2017}.
But the first evidence for exocomets transiting in front of a star in
visible light had to wait until August 2017.
It comes from the discovery of possibly several comets around the star
KIC~3542116~\citep{rappaport2018}.
This pioneering paper further analyses the possible properties of the
comets based on the shape of the observed light curves, which are book
examples of the expectations in~\citep{lecavelier1999}.
The authors of the paper are optimist that new examples will be found
in future analyses of the \emph{Kepler} data.

One disputed case is KIC~8462852~\citep{boyajian2016}, a
target observed by the \emph{Kepler} mission that show irregular flux
drops that account for up to 20\% of the stellar flux lasting several
days.
Some teams have invoked the possibility of comets~\citep{bodman2016}
to explain the observations, but the hypothesis has been recently
challenged~\citep{wright2016}.
However, new observations in May 2017 (triggered by a \emph{tweet} by 
T.~Boyajian,~\url{@tsboyajian}) suggest that the phenomenon has a
characteristic time-scale of 700 days and its origin, whatever its
nature, is indeed gravitationally linked to the star.
It has been proposed that ringed planet and a large swarm of trojan
bodies could explain the features observed in the light curve and
predict its future behaviour~\citep{ballesteros2017}.
The semi-major axis of the system would be around 6 au and the orbital
period 12 years, a large observational span, but certainly within
reach in the near future.

\section{Trojans}

In the solar system the term Trojans refer to a family of minor bodies
that share the orbits of the giant planets like Jupiter and Neptune.
They represent a special case of the co-orbital dynamics of the N-body
problem~\citep[see, for example,][]{veras2016}.

The stability of such configurations is not simple, nor their
dynamical evolution as planets migrate~\citep{nesvorny2013b}, but
there is no known reason preventing their existence in extrasolar
systems.
Only that, if they have similar sizes to those in the solar system,
their direct detectability with photometry is beyond reach for current
and near future facilities.

Therefore, researchers have tried to infer the presence of trojans
studying the perturbations introduce in the orbit of their larger,
companion planet, both with radial velocity, transit timing
variations, or
both~\citep{ford2007,dobrovolskis2013,haghighipour2013,leleu2015,leleu2017,nesvorny2016,vokrouhlick2014}.
There have been several systematic searches that have not found any
reliable candidate so far~\citep{madhusudhan2008,janson2013}.

There have been claims in the literature with detections, but so far
none of these claims has been confirmed by subsequent independent
analysis, like the cases of HD~82943 and
HD~128311~\citep{gozdziewski2006}, but see~\citet{mcarthur2014}
and~\citet{rein2015}, or the Kepler candidates by~\citet{hippke2015b},
which at least in the case of Kepler-91 b have not been confirmed by
later studies~\citep{placek2015}.
A recent claim on WASP-12 and HD~18733 by~\citet{kislyakova2016}
remains to be confirmed.

In this case, the direct confirmation of the presence of a trojan
suffers from the same limitations of reproducibility, credibility in
presence of correlated noise, and degeneracy of interpretation as in
the previous examples of moons and rings, with the detriment of the
smaller size and mass of the researched object.

\section{Summary}

We would like to close with a quote attributed to Napoleon Bonaparte
dissuading from taking predictions too seriously
(\emph{Il faut toujours se r{\'e}server le droit de rire le lendemain
  de ses id{\'e}es de la veille}).
Or in the words of a famous astronomer, \emph{you should listen to
theorists but never take them too seriously}.
The discovery and characterization of moons, rings, comets, and
trojans has proven more challenging than expected, but the interest of
the community has not decreased in the last 20 years.
It is rather the opposite.
And there are also new ideas coming out, like
\emph{synestias}~\citep{lock2017}.
These are transient structures predicted by theoretical models
produced during planetary formation processes.
They have not been observed, or confirmed independently yet, but they
are certainly welcome because of their interest.

The wealth of data from transit photometry has been carefully studied
and most of the systematics are well understood, yet no undisputed
detection has been definitely accepted by the community. 
However, the situation is quickly changing in a very positive way.
TESS and PLATO will have their chance in the next decade, but it is
time to think about different methods, in particular direct 
imaging and probably high resolution spectroscopy (for example, the 
serendipitous discovery of a moon or ring system via
Rossiter-McLaughlin during planet characterization).
With instruments like Gaia, ALMA, and E-ELT class telescopes it is
difficult not to end with the conclusion that moons, rings, comets,
and trojans will not only be detected in large numbers in the next
decades, but also will contribute to our knowledge about planets and
planetary systems, in our galaxy and in the solar system.

\section{Cross-References}
\begin{itemize}
\item{On the Detection of Extrasolar Moons and Rings}
\end{itemize}


\bibliographystyle{spbasicHBexo}  
\bibliography{bibl} 

\begin{thebibliography}{149}
\providecommand{\natexlab}[1]{#1}
\providecommand{\url}[1]{{#1}}
\providecommand{\urlprefix}{URL }
\expandafter\ifx\csname urlstyle\endcsname\relax
  \providecommand{\doi}[1]{DOI~\discretionary{}{}{}#1}\else
  \providecommand{\doi}{DOI~\discretionary{}{}{}\begingroup
  \urlstyle{rm}\Url}\fi
\providecommand{\eprint}[2][]{\url{#2}}

\bibitem[{{Adams} and {Bloch}(2016)}]{adams2016b}
{Adams} FC {Bloch} AM (2016) {The stability of tidal equilibrium for
  hierarchical star-planet-moon systems}. \mnras 462:2527--2541

\bibitem[{{Aizawa} et~al.(2017){Aizawa}, {Uehara}, {Masuda}, {Kawahara}, and
  {Suto}}]{aizawa2017}
{Aizawa} M, {Uehara} S, {Masuda} K, {Kawahara} H {Suto} Y (2017) {Toward
  Detection of Exoplanetary Rings via Transit Photometry: Methodology and a
  Possible Candidate}. \aj 153:193

\bibitem[{{Alonso} et~al.(2016){Alonso}, {Rappaport}, {Deeg}, and
  {Palle}}]{alonso2016}
{Alonso} R, {Rappaport} S, {Deeg} HJ {Palle} E (2016) {Gray transits of WD
  1145+017 over the visible band}. \aap 589:L6

\bibitem[{{Arnold} and {Schneider}(2004)}]{arnold2004}
{Arnold} L {Schneider} J (2004) {The detectability of extrasolar planet
  surroundings. I. Reflected-light photometry of unresolved rings}. \aap
  420:1153--1162

\bibitem[{{Ballesteros} et~al.(2018){Ballesteros}, {Arnalte-Mur},
  {Fernandez-Soto}, and {Mart{\'{\i}}nez}}]{ballesteros2017}
{Ballesteros} FJ, {Arnalte-Mur} P, {Fernandez-Soto} A {Mart{\'{\i}}nez} VJ
  (2018) {KIC 8462852: Will the Trojans return in 2021?} \mnras 473:L21--L25

\bibitem[{{Barnes} and {O'Brien}(2002)}]{barnes2002}
{Barnes} JW {O'Brien} DP (2002) {Stability of Satellites around Close-in
  Extrasolar Giant Planets}. \apj 575:1087--1093

\bibitem[{Barr(2016)}]{barr2017a}
Barr AC (2016) Formation of exomoons: a solar system perspective. Astronomical
  Review 12(1-4):24--52

\bibitem[{{Barr} and {Bruck Syal}(2017)}]{barr2017b}
{Barr} AC {Bruck Syal} M (2017) {Formation of massive rocky exomoons by giant
  impact}. \mnras 466:4868--4874

\bibitem[{{Barros} et~al.(2013){Barros}, {Bou{\'e}}, {Gibson}, {Pollacco},
  {Santerne}, {Keenan}, {Skillen}, and {Street}}]{barros2013}
{Barros} SCC, {Bou{\'e}} G, {Gibson} NP et~al. (2013) {Transit timing
  variations in WASP-10b induced by stellar activity}. \mnras 430:3032--3047

\bibitem[{{Bedell} et~al.(2015){Bedell}, {Mel{\'e}ndez}, {Bean},
  {Ram{\'{\i}}rez}, {Asplund}, {Alves-Brito}, {Casagrande}, {Dreizler},
  {Monroe}, {Spina}, and {Tucci Maia}}]{bedell2015}
{Bedell} M, {Mel{\'e}ndez} J, {Bean} JL et~al. (2015) {The Solar Twin Planet
  Search. II. A Jupiter twin around a solar twin}. \aap 581:A34

\bibitem[{{Bennett} et~al.(2014){Bennett}, {Batista}, {Bond}, {Bennett},
  {Suzuki}, {Beaulieu}, {Udalski}, {Donatowicz}, {Bozza}, {Abe}, {Botzler},
  {Freeman}, {Fukunaga}, {Fukui}, {Itow}, {Koshimoto}, {Ling}, {Masuda},
  {Matsubara}, {Muraki}, {Namba}, {Ohnishi}, {Rattenbury}, {Saito}, {Sullivan},
  {Sumi}, {Sweatman}, {Tristram}, {Tsurumi}, {Wada}, {Yock}, {MOA
  Collaboration}, {Albrow}, {Bachelet}, {Brillant}, {Caldwell}, {Cassan},
  {Cole}, {Corrales}, {Coutures}, {Dieters}, {Dominis Prester}, {Fouqu{\'e}},
  {Greenhill}, {Horne}, {Koo}, {Kubas}, {Marquette}, {Martin}, {Menzies},
  {Sahu}, {Wambsganss}, {Williams}, {Zub}, {PLANET Collaboration}, {Choi},
  {DePoy}, {Dong}, {Gaudi}, {Gould}, {Han}, {Henderson}, {McGregor}, {Lee},
  {Pogge}, {Shin}, {Yee}, {{$\mu$}FUN Collaboration}, {Szyma{\'n}ski},
  {Skowron}, {Poleski}, {Koz{\l}owski}, {Wyrzykowski}, {Kubiak},
  {Pietrukowicz}, {Pietrzy{\'n}ski}, {Soszy{\'n}ski}, {Ulaczyk}, {OGLE
  Collaboration}, {Tsapras}, {Street}, {Dominik}, {Bramich}, {Browne},
  {Hundertmark}, {Kains}, {Snodgrass}, {Steele}, {RoboNet Collaboration},
  {Dekany}, {Gonzalez}, {Heyrovsk{\'y}}, {Kandori}, {Kerins}, {Lucas},
  {Minniti}, {Nagayama}, {Rejkuba}, {Robin}, and {Saito}}]{bennet2014}
{Bennett} DP, {Batista} V, {Bond} IA et~al. (2014) {MOA-2011-BLG-262Lb: A
  Sub-Earth-Mass Moon Orbiting a Gas Giant Primary or a High Velocity Planetary
  System in the Galactic Bulge}. \apj 785:155

\bibitem[{{Best} et~al.(2017){Best}, {Liu}, {Dupuy}, and {Magnier}}]{best2017}
{Best} WMJ, {Liu} MC, {Dupuy} TJ {Magnier} EA (2017) {The Young L Dwarf 2MASS
  J11193254-1137466 Is a Planetary-mass Binary}. \apjl 843:L4

\bibitem[{{Bodman} and {Quillen}(2016)}]{bodman2016}
{Bodman} EHL {Quillen} A (2016) {KIC 8462852: Transit of a Large Comet Family}.
  \apjl 819:L34

\bibitem[{{Bowler}(2016)}]{bowler2016}
{Bowler} BP (2016) {Imaging Extrasolar Giant Planets}. \pasp 128(10):102,001

\bibitem[{{Boyajian} et~al.(2016){Boyajian}, {LaCourse}, {Rappaport},
  {Fabrycky}, {Fischer}, {Gandolfi}, {Kennedy}, {Korhonen}, {Liu}, {Moor},
  {Olah}, {Vida}, {Wyatt}, {Best}, {Brewer}, {Ciesla}, {Cs{\'a}k}, {Deeg},
  {Dupuy}, {Handler}, {Heng}, {Howell}, {Ishikawa}, {Kov{\'a}cs}, {Kozakis},
  {Kriskovics}, {Lehtinen}, {Lintott}, {Lynn}, {Nespral}, {Nikbakhsh},
  {Schawinski}, {Schmitt}, {Smith}, {Szabo}, {Szabo}, {Viuho}, {Wang},
  {Weiksnar}, {Bosch}, {Connors}, {Goodman}, {Green}, {Hoekstra}, {Jebson},
  {Jek}, {Omohundro}, {Schwengeler}, and {Szewczyk}}]{boyajian2016}
{Boyajian} TS, {LaCourse} DM, {Rappaport} SA et~al. (2016) {Planet Hunters IX.
  KIC 8462852 - where's the flux?} \mnras 457:3988--4004

\bibitem[{{Brown} et~al.(2001){Brown}, {Charbonneau}, {Gilliland}, {Noyes}, and
  {Burrows}}]{brown2001}
{Brown} TM, {Charbonneau} D, {Gilliland} RL, {Noyes} RW {Burrows} A (2001)
  {Hubble Space Telescope Time-Series Photometry of the Transiting Planet of HD
  209458}. \apj 552:699--709

\bibitem[{{Cabrera} and {Schneider}(2007)}]{cabrera2007}
{Cabrera} J {Schneider} J (2007) {Detecting companions to extrasolar planets
  using mutual events}. \aap 464:1133--1138

\bibitem[{{Casertano} et~al.(2008){Casertano}, {Lattanzi}, {Sozzetti},
  {Spagna}, {Jancart}, {Morbidelli}, {Pannunzio}, {Pourbaix}, and
  {Queloz}}]{casertano2008}
{Casertano} S, {Lattanzi} MG, {Sozzetti} A et~al. (2008) {Double-blind test
  program for astrometric planet detection with Gaia}. \aap 482:699--729

\bibitem[{{Cassidy} et~al.(2009){Cassidy}, {Mendez}, {Arras}, {Johnson}, and
  {Skrutskie}}]{cassidy2009}
{Cassidy} TA, {Mendez} R, {Arras} P, {Johnson} RE {Skrutskie} MF (2009)
  {Massive Satellites of Close-In Gas Giant Exoplanets}. \apj 704:1341--1348

\bibitem[{{Cockell} et~al.(2016){Cockell}, {Bush}, {Bryce}, {Direito},
  {Fox-Powell}, {Harrison}, {Lammer}, {Landenmark}, {Martin-Torres},
  {Nicholson}, {Noack}, {O'Malley-James}, {Payler}, {Rushby}, {Samuels},
  {Schwendner}, {Wadsworth}, and {Zorzano}}]{cockell2016}
{Cockell} CS, {Bush} T, {Bryce} C et~al. (2016) {Habitability: A Review}.
  Astrobiology 16:89--117

\bibitem[{{Crida} and {Charnoz}(2012)}]{crida2012}
{Crida} A {Charnoz} S (2012) {Formation of Regular Satellites from Ancient
  Massive Rings in the Solar System}. Science 338:1196--

\bibitem[{{David} et~al.(2017){David}, {Petigura}, {Hillenbrand}, {Cody},
  {Collier Cameron}, {Stauffer}, {Fulton}, {Isaacson}, {Howard}, {Howell},
  {Everett}, {Wang}, {Benneke}, {Hellier}, {West}, {Pollacco}, and
  {Anderson}}]{david2017}
{David} TJ, {Petigura} EA, {Hillenbrand} LA et~al. (2017) {A Transient Transit
  Signature Associated with the Young Star RIK-210}. \apj 835:168

\bibitem[{{de Pater} and {Lissauer}(2015)}]{depater2015}
{de Pater} I {Lissauer} JJ (2015) {Planetary Sciences}

\bibitem[{{Debes} and {Sigurdsson}(2007)}]{debes2007}
{Debes} JH {Sigurdsson} S (2007) {The Survival Rate of Ejected Terrestrial
  Planets with Moons}. \apjl 668:L167--L170

\bibitem[{{D{\'{\i}}az} et~al.(2016){D{\'{\i}}az}, {Rey}, {Demangeon},
  {H{\'e}brard}, {Boisse}, {Arnold}, {Astudillo-Defru}, {Beuzit}, {Bonfils},
  {Borgniet}, {Bouchy}, {Bourrier}, {Courcol}, {Deleuil}, {Delfosse},
  {Ehrenreich}, {Forveille}, {Lagrange}, {Mayor}, {Moutou}, {Pepe}, {Queloz},
  {Santerne}, {Santos}, {Sahlmann}, {S{\'e}gransan}, {Udry}, and
  {Wilson}}]{diaz2016b}
{D{\'{\i}}az} RF, {Rey} J, {Demangeon} O et~al. (2016) {The SOPHIE search for
  northern extrasolar planets. XI. Three new companions and an orbit update:
  Giant planets in the habitable zone}. \aap 591:A146

\bibitem[{{Dobos} et~al.(2016){Dobos}, {Kereszturi}, {P{\'a}l}, and
  {Kiss}}]{dobos2016}
{Dobos} V, {Kereszturi} {\'A}, {P{\'a}l} A {Kiss} LL (2016) {Possibility for
  albedo estimation of exomoons: Why should we care about M dwarfs?} \aap
  592:A139

\bibitem[{{Dobos} et~al.(2017){Dobos}, {Heller}, and {Turner}}]{dobos2017}
{Dobos} V, {Heller} R {Turner} EL (2017) {The effect of multiple heat sources
  on exomoon habitable zones}. \aap 601:A91

\bibitem[{{Dobrovolskis}(2013)}]{dobrovolskis2013}
{Dobrovolskis} AR (2013) {Effects of Trojan exoplanets on the reflex motions of
  their parent stars}. \icarus 226:1635--1641

\bibitem[{{Domingos} et~al.(2006){Domingos}, {Winter}, and
  {Yokoyama}}]{domingos2006}
{Domingos} RC, {Winter} OC {Yokoyama} T (2006) {Stable satellites around
  extrasolar giant planets}. \mnras 373:1227--1234

\bibitem[{{Donnison}(2010)}]{donnison2010}
{Donnison} JR (2010) {The Hill stability of the possible moons of extrasolar
  planets}. \mnras 406:1918--1934

\bibitem[{{Ehrenreich} et~al.(2012){Ehrenreich}, {Bourrier}, {Bonfils},
  {Lecavelier des Etangs}, {H{\'e}brard}, {Sing}, {Wheatley}, {Vidal-Madjar},
  {Delfosse}, {Udry}, {Forveille}, and {Moutou}}]{ehrenreich2012}
{Ehrenreich} D, {Bourrier} V, {Bonfils} X et~al. (2012) {Hint of a transiting
  extended atmosphere on 55 Cancri b}. \aap 547:A18

\bibitem[{{Ehrenreich} et~al.(2015){Ehrenreich}, {Bourrier}, {Wheatley},
  {Lecavelier des Etangs}, {H{\'e}brard}, {Udry}, {Bonfils}, {Delfosse},
  {D{\'e}sert}, {Sing}, and {Vidal-Madjar}}]{ehrenreich2015}
{Ehrenreich} D, {Bourrier} V, {Wheatley} PJ et~al. (2015) {A giant comet-like
  cloud of hydrogen escaping the warm Neptune-mass exoplanet GJ 436b}. \nat
  522:459--461

\bibitem[{{Eiroa} et~al.(2016){Eiroa}, {Rebollido}, {Montesinos}, {Villaver},
  {Absil}, {Henning}, {Bayo}, {Canovas}, {Carmona}, {Chen}, {Ertel},
  {Iglesias}, {Launhardt}, {Maldonado}, {Meeus}, {Mo{\'o}r}, {Mora}, {Mustill},
  {Olofsson}, {Riviere-Marichalar}, and {Roberge}}]{eiroa2016}
{Eiroa} C, {Rebollido} I, {Montesinos} B et~al. (2016) {Exocomet signatures
  around the A-shell star phis Leonis?} \aap 594:L1

\bibitem[{{Esposito} et~al.(2013){Esposito}, {Mesa}, {Skemer}, {Arcidiacono},
  {Claudi}, {Desidera}, {Gratton}, {Mannucci}, {Marzari}, {Masciadri}, {Close},
  {Hinz}, {Kulesa}, {McCarthy}, {Males}, {Agapito}, {Argomedo}, {Boutsia},
  {Briguglio}, {Brusa}, {Busoni}, {Cresci}, {Fini}, {Fontana}, {Guerra},
  {Hill}, {Miller}, {Paris}, {Pinna}, {Puglisi}, {Quiros-Pacheco}, {Riccardi},
  {Stefanini}, {Testa}, {Xompero}, and {Woodward}}]{esposito2013}
{Esposito} S, {Mesa} D, {Skemer} A et~al. (2013) {LBT observations of the HR
  8799 planetary system. First detection of HR 8799e in H band}. \aap 549:A52

\bibitem[{{Ford} and {Holman}(2007)}]{ford2007}
{Ford} EB {Holman} MJ (2007) {Using Transit Timing Observations to Search for
  Trojans of Transiting Extrasolar Planets}. \apjl 664:L51--L54

\bibitem[{{Foreman-Mackey} et~al.(2016){Foreman-Mackey}, {Morton}, {Hogg},
  {Agol}, and {Sch{\"o}lkopf}}]{foremanmackey2016}
{Foreman-Mackey} D, {Morton} TD, {Hogg} DW, {Agol} E {Sch{\"o}lkopf} B (2016)
  {The Population of Long-period Transiting Exoplanets}. \aj 152:206

\bibitem[{{Forgan} and {Kipping}(2013)}]{forgan2013}
{Forgan} D {Kipping} D (2013) {Dynamical effects on the habitable zone for
  Earth-like exomoons}. \mnras 432:2994--3004

\bibitem[{{Forgan} and {Yotov}(2014)}]{forgan2014}
{Forgan} D {Yotov} V (2014) {The effect of planetary illumination on climate
  modelling of Earth-like exomoons}. \mnras 441:3513--3523

\bibitem[{{Fujii} et~al.(2017){Fujii}, {Kobayashi}, {Takahashi}, and
  {Gressel}}]{fujii2017}
{Fujii} YI, {Kobayashi} H, {Takahashi} SZ {Gressel} O (2017) {Orbital Evolution
  of Moons in Weakly Accreting Circumplanetary Disks}. \aj 153:194

\bibitem[{{Gaia Collaboration} et~al.(2016){Gaia Collaboration}, {Brown},
  {Vallenari}, {Prusti}, {de Bruijne}, {Mignard}, {Drimmel}, {Babusiaux},
  {Bailer-Jones}, {Bastian}, and et~al.}]{brown2016GaiaDR1}
{Gaia Collaboration}, {Brown} AGA, {Vallenari} A et~al. (2016) {Gaia Data
  Release 1. Summary of the astrometric, photometric, and survey properties}.
  \aap 595:A2

\bibitem[{{Gaidos} et~al.(2016){Gaidos}, {Mann}, and {Ansdell}}]{gaidos2016}
{Gaidos} E, {Mann} AW {Ansdell} M (2016) {The Enigmatic and Ephemeral M Dwarf
  System KOI 6705: Cheshire Cat or Wild Goose?} \apj 817:50

\bibitem[{{Gaudi} et~al.(2003){Gaudi}, {Chang}, and {Han}}]{gaudi2003}
{Gaudi} BS, {Chang} HY {Han} C (2003) {Probing Structures of Distant Extrasolar
  Planets with Microlensing}. \apj 586:527--539

\bibitem[{{Gong} et~al.(2013){Gong}, {Zhou}, {Xie}, and {Wu}}]{gong2013}
{Gong} YX, {Zhou} JL, {Xie} JW {Wu} XM (2013) {The Effect of Planet-Planet
  Scattering on the Survival of Exomoons}. \apjl 769:L14

\bibitem[{{Go{\'z}dziewski} and {Konacki}(2006)}]{gozdziewski2006}
{Go{\'z}dziewski} K {Konacki} M (2006) {Trojan Pairs in the HD 128311 and HD
  82943 Planetary Systems?} \apj 647:573--586

\bibitem[{{Greco} and {Burrows}(2015)}]{greco2015}
{Greco} JP {Burrows} A (2015) {The Direct Detectability of Giant Exoplanets in
  the Optical}. \apj 808:172

\bibitem[{{Haghighipour} et~al.(2013){Haghighipour}, {Capen}, and
  {Hinse}}]{haghighipour2013}
{Haghighipour} N, {Capen} S {Hinse} TC (2013) {Detection of Earth-mass and
  super-Earth Trojan planets using transit timing variation method}. Celestial
  Mechanics and Dynamical Astronomy 117:75--89

\bibitem[{{Han} et~al.(2017){Han}, {Udalski}, {Sumi}, {Gould}, {Albrow},
  {Chung}, {Jung}, {Ryu}, {Shin}, {Yee}, {Zhu}, {Cha}, {Kim}, {Kim}, {Lee},
  {Lee}, {Park}, {KMTNet Collaboration}, {Soszy{\'n}ski}, {Mr{\'o}z},
  {Pietrukowicz}, {Szyma{\'n}ski}, {Skowron}, {Poleski}, {Koz{\l}owski},
  {Ulaczyk}, {Pawlak}, {The OGLE Collaboration}, {Abe}, {Asakura}, {Bennett},
  {Bond}, {Bhattacharya}, {Donachie}, {Freeman}, {Fukui}, {Hirao}, {Itow},
  {Koshimoto}, {Li}, {Ling}, {Masuda}, {Matsubara}, {Muraki}, {Nagakane},
  {Ohnishi}, {Oyokawa}, {Rattenbury}, {Saito}, {Sharan}, {Sullivan}, {Suzuki},
  {Tristram}, {Yamada}, {Yamada}, {Yonehara}, {Barry}, and {The MOA
  Collaboration}}]{han2017}
{Han} C, {Udalski} A, {Sumi} T et~al. (2017) {OGLE-2016-BLG-1469L: Microlensing
  Binary Composed of Brown Dwarfs}. \apj 843:59

\bibitem[{{Haswell} et~al.(2012){Haswell}, {Fossati}, {Ayres}, {France},
  {Froning}, {Holmes}, {Kolb}, {Busuttil}, {Street}, {Hebb}, {Collier Cameron},
  {Enoch}, {Burwitz}, {Rodriguez}, {West}, {Pollacco}, {Wheatley}, and
  {Carter}}]{haswell2012}
{Haswell} CA, {Fossati} L, {Ayres} T et~al. (2012) {Near-ultraviolet
  Absorption, Chromospheric Activity, and Star-Planet Interactions in the
  WASP-12 system}. \apj 760:79

\bibitem[{{Heising} et~al.(2015){Heising}, {Marcy}, and
  {Schlichting}}]{heising2015}
{Heising} MZ, {Marcy} GW {Schlichting} HE (2015) {A Search for Ringed
  Exoplanets Using Kepler Photometry}. \apj 814:81

\bibitem[{{Heller}(2012)}]{heller2012}
{Heller} R (2012) {Exomoon habitability constrained by energy flux and orbital
  stability}. \aap 545:L8

\bibitem[{{Heller} and {Barnes}(2015)}]{heller2015b}
{Heller} R {Barnes} R (2015) {Runaway greenhouse effect on exomoons due to
  irradiation from hot, young giant planets}. International Journal of
  Astrobiology 14:335--343

\bibitem[{{Heller} and {Pudritz}(2015)}]{heller2015a}
{Heller} R {Pudritz} R (2015) {Conditions for water ice lines and Mars-mass
  exomoons around accreting super-Jovian planets at 1-20 AU from Sun-like
  stars}. \aap 578:A19

\bibitem[{{Hippke}(2015)}]{hippke2015a}
{Hippke} M (2015) {On the Detection of Exomoons: A Search in Kepler Data for
  the Orbital Sampling Effect and the Scatter Peak}. \apj 806:51

\bibitem[{{Hippke} and {Angerhausen}(2015)}]{hippke2015b}
{Hippke} M {Angerhausen} D (2015) {A Statistical Search for a Population of
  Exo-Trojans in the Kepler Data Set}. \apj 811:1

\bibitem[{{Hong} et~al.(2015){Hong}, {Tiscareno}, {Nicholson}, and
  {Lunine}}]{hong2015}
{Hong} YC, {Tiscareno} MS, {Nicholson} PD {Lunine} JI (2015) {Orbital
  instability of close-in exomoons in non-coplanar systems}. \mnras
  449:828--834

\bibitem[{{Janson}(2013)}]{janson2013}
{Janson} M (2013) {A Systematic Search for Trojan Planets in the Kepler Data}.
  \apj 774:156

\bibitem[{{Kaltenegger}(2010)}]{kaltenegger2010a}
{Kaltenegger} L (2010) {Characterizing Habitable Exomoons}. \apjl
  712:L125--L130

\bibitem[{{Kane}(2017)}]{kane2017}
{Kane} SR (2017) {Worlds without Moons: Exomoon Constraints for Compact
  Planetary Systems}. \apjl 839:L19

\bibitem[{{Karkoschka}(1994)}]{karkoschka1994}
{Karkoschka} E (1994) {Spectrophotometry of the jovian planets and Titan at
  300- to 1000-nm wavelength: The methane spectrum}. \icarus 111:174--192

\bibitem[{{Kenworthy}(2017)}]{kenworthy2017}
{Kenworthy} M (2017) {Looking for rings and things}. Nature Astronomy 1:0099

\bibitem[{{Kiefer} et~al.(2014{\natexlab{a}}){Kiefer}, {Lecavelier des Etangs},
  {Augereau}, {Vidal-Madjar}, {Lagrange}, and {Beust}}]{kiefer2014b}
{Kiefer} F, {Lecavelier des Etangs} A, {Augereau} JC et~al.
  (2014{\natexlab{a}}) {Exocomets in the circumstellar gas disk of HD 172555}.
  \aap 561:L10

\bibitem[{{Kiefer} et~al.(2014{\natexlab{b}}){Kiefer}, {Lecavelier des Etangs},
  {Boissier}, {Vidal-Madjar}, {Beust}, {Lagrange}, {H{\'e}brard}, and
  {Ferlet}}]{kiefer2014a}
{Kiefer} F, {Lecavelier des Etangs} A, {Boissier} J et~al. (2014{\natexlab{b}})
  {Two families of exocomets in the {$\beta$} Pictoris system}. \nat
  514:462--464

\bibitem[{{Kipping} et~al.(2012){Kipping}, {Bakos}, {Buchhave}, {Nesvorn{\'y}},
  and {Schmitt}}]{kipping2012c}
{Kipping} DM, {Bakos} G{\'A}, {Buchhave} L, {Nesvorn{\'y}} D {Schmitt} A (2012)
  {The Hunt for Exomoons with Kepler (HEK). I. Description of a New
  Observational project}. \apj 750:115

\bibitem[{{Kipping} et~al.(2013{\natexlab{a}}){Kipping}, {Forgan}, {Hartman},
  {Nesvorn{\'y}}, {Bakos}, {Schmitt}, and {Buchhave}}]{kipping2013b}
{Kipping} DM, {Forgan} D, {Hartman} J et~al. (2013{\natexlab{a}}) {The Hunt for
  Exomoons with Kepler (HEK). III. The First Search for an Exomoon around a
  Habitable-zone Planet}. \apj 777:134

\bibitem[{{Kipping} et~al.(2013{\natexlab{b}}){Kipping}, {Hartman}, {Buchhave},
  {Schmitt}, {Bakos}, and {Nesvorn{\'y}}}]{kipping2013a}
{Kipping} DM, {Hartman} J, {Buchhave} LA et~al. (2013{\natexlab{b}}) {The Hunt
  for Exomoons with Kepler (HEK). II. Analysis of Seven Viable
  Satellite-hosting Planet Candidates}. \apj 770:101

\bibitem[{{Kipping} et~al.(2014{\natexlab{a}}){Kipping}, {Nesvorn{\'y}},
  {Buchhave}, {Hartman}, {Bakos}, and {Schmitt}}]{kipping2014b}
{Kipping} DM, {Nesvorn{\'y}} D, {Buchhave} LA et~al. (2014{\natexlab{a}}) {The
  Hunt for Exomoons with Kepler (HEK). IV. A Search for Moons around Eight M
  Dwarfs}. \apj 784:28

\bibitem[{{Kipping} et~al.(2014{\natexlab{b}}){Kipping}, {Torres}, {Buchhave},
  {Kenyon}, {Henze}, {Isaacson}, {Kolbl}, {Marcy}, {Bryson}, {Stassun}, and
  {Bastien}}]{kipping2014c}
{Kipping} DM, {Torres} G, {Buchhave} LA et~al. (2014{\natexlab{b}}) {Discovery
  of a Transiting Planet near the Snow-line}. \apj 795:25

\bibitem[{{Kipping} et~al.(2015{\natexlab{a}}){Kipping}, {Huang},
  {Nesvorn{\'y}}, {Torres}, {Buchhave}, {Bakos}, and {Schmitt}}]{kipping2015a}
{Kipping} DM, {Huang} X, {Nesvorn{\'y}} D et~al. (2015{\natexlab{a}}) {The
  Possible Moon of Kepler-90g is a False Positive}. \apjl 799:L14

\bibitem[{{Kipping} et~al.(2015{\natexlab{b}}){Kipping}, {Schmitt}, {Huang},
  {Torres}, {Nesvorn{\'y}}, {Buchhave}, {Hartman}, and {Bakos}}]{kipping2015b}
{Kipping} DM, {Schmitt} AR, {Huang} X et~al. (2015{\natexlab{b}}) {The Hunt for
  Exomoons with Kepler (HEK): V. A Survey of 41 Planetary Candidates for
  Exomoons}. \apj 813:14

\bibitem[{{Kipping} et~al.(2016){Kipping}, {Torres}, {Henze}, {Teachey},
  {Isaacson}, {Petigura}, {Marcy}, {Buchhave}, {Chen}, {Bryson}, and
  {Sandford}}]{kipping2016b}
{Kipping} DM, {Torres} G, {Henze} C et~al. (2016) {A Transiting Jupiter
  Analog}. \apj 820:112

\bibitem[{{Kislyakova} et~al.(2016){Kislyakova}, {Pilat-Lohinger}, {Funk},
  {Lammer}, {Fossati}, {Eggl}, {Schwarz}, {Boudjada}, and
  {Erkaev}}]{kislyakova2016}
{Kislyakova} KG, {Pilat-Lohinger} E, {Funk} B et~al. (2016) {On the ultraviolet
  anomalies of the WASP-12 and HD 189733 systems: Trojan satellites as a plasma
  source}. \mnras 461:988--999

\bibitem[{{Lagrange} et~al.(2010){Lagrange}, {Bonnefoy}, {Chauvin}, {Apai},
  {Ehrenreich}, {Boccaletti}, {Gratadour}, {Rouan}, {Mouillet}, {Lacour}, and
  {Kasper}}]{lagrange2010}
{Lagrange} AM, {Bonnefoy} M, {Chauvin} G et~al. (2010) {A Giant Planet Imaged
  in the Disk of the Young Star {$\beta$} Pictoris}. Science 329:57

\bibitem[{{Lammer} et~al.(2014){Lammer}, {Schiefer}, {Juvan}, {Odert},
  {Erkaev}, {Weber}, {Kislyakova}, {G{\"u}del}, {Kirchengast}, and
  {Hanslmeier}}]{lammer2014}
{Lammer} H, {Schiefer} SC, {Juvan} I et~al. (2014) {Origin and Stability of
  Exomoon Atmospheres: Implications for Habitability}. Origins of Life and
  Evolution of the Biosphere 44:239--260

\bibitem[{{Latter} et~al.(2017){Latter}, {Ogilvie}, and {Rein}}]{latter2017}
{Latter} HN, {Ogilvie} GI {Rein} H (2017) Planetary rings and other
  astrophysical disks. In: Planetary Ring Systems

\bibitem[{{Laughlin} and {Adams}(2000)}]{laughlin2000}
{Laughlin} G {Adams} FC (2000) {The Frozen Earth: Binary Scattering Events and
  the Fate of the Solar System}. \icarus 145:614--627

\bibitem[{{Lecavelier des Etangs} and {Vidal-Madjar}(2016)}]{lecavelier2016}
{Lecavelier des Etangs} A {Vidal-Madjar} A (2016) {The orbit of beta Pictoris b
  as a transiting planet}. \aap 588:A60

\bibitem[{{Lecavelier Des Etangs} et~al.(1999){Lecavelier Des Etangs},
  {Vidal-Madjar}, and {Ferlet}}]{lecavelier1999}
{Lecavelier Des Etangs} A, {Vidal-Madjar} A {Ferlet} R (1999) {Photometric
  stellar variation due to extra-solar comets}. \aap 343:916--922

\bibitem[{{Lecavelier Des Etangs} et~al.(2010){Lecavelier Des Etangs},
  {Ehrenreich}, {Vidal-Madjar}, {Ballester}, {D{\'e}sert}, {Ferlet},
  {H{\'e}brard}, {Sing}, {Tchakoumegni}, and {Udry}}]{lecavelier2010}
{Lecavelier Des Etangs} A, {Ehrenreich} D, {Vidal-Madjar} A et~al. (2010)
  {Evaporation of the planet HD 189733b observed in H I Lyman-{$\alpha$}}. \aap
  514:A72

\bibitem[{{Lecavelier des Etangs} et~al.(2017){Lecavelier des Etangs},
  {H{\'e}brard}, {Blandin}, {Cassier}, {Deeg}, {Bonomo}, {Bouchy},
  {D{\'e}sert}, {Ehrenreich}, {Deleuil}, {D{\'{\i}}az}, {Moutou}, and
  {Vidal-Madjar}}]{lecavelier2017}
{Lecavelier des Etangs} A, {H{\'e}brard} G, {Blandin} S et~al. (2017) {Search
  for rings and satellites around the exoplanet CoRoT-9b using Spitzer
  photometry}. \aap 603:A115

\bibitem[{{Leleu} et~al.(2015){Leleu}, {Robutel}, and {Correia}}]{leleu2015}
{Leleu} A, {Robutel} P {Correia} ACM (2015) {Detectability of quasi-circular
  co-orbital planets. Application to the radial velocity technique}. \aap
  581:A128

\bibitem[{{Leleu} et~al.(2017){Leleu}, {Robutel}, {Correia}, and
  {Lillo-Box}}]{leleu2017}
{Leleu} A, {Robutel} P, {Correia} ACM {Lillo-Box} J (2017) {Detection of
  co-orbital planets by combining transit and radial-velocity measurements}.
  \aap 599:L7

\bibitem[{{Lester} et~al.(1979){Lester}, {McCall}, and {Tatum}}]{lester1979}
{Lester} TP, {McCall} ML {Tatum} JB (1979) {Theory of planetary photometry}.
  \jrasc 73:233--257

\bibitem[{{Lewis}(2013)}]{lewis2013}
{Lewis} KM (2013) {Detecting exo-moons with photometric transit timing - I.
  Effect of realistic stellar noise}. \mnras 430:1473--1485

\bibitem[{{Lewis} and {Fujii}(2014)}]{lewis2014}
{Lewis} KM {Fujii} Y (2014) {Next Generation of Telescopes or Dynamics Required
  to Determine if Exo-moons have Prograde or Retrograde Orbits}. \apjl 791:L26

\bibitem[{{Lewis} et~al.(2015){Lewis}, {Ochiai}, {Nagasawa}, and
  {Ida}}]{lewis2015}
{Lewis} KM, {Ochiai} H, {Nagasawa} M {Ida} S (2015) {Extrasolar Binary Planets
  II: Detectability by Transit Observations}. \apj 805:27

\bibitem[{{Li} et~al.(2016){Li}, {Tian}, {Wang}, {Wei}, and {Huang}}]{li2016}
{Li} T, {Tian} F, {Wang} Y, {Wei} W {Huang} X (2016) {Distinguishing a
  Hypothetical Abiotic Planet-Moon System from a Single Inhabited Planet}.
  \apjl 817:L15

\bibitem[{{Lock} and {Stewart}(2017)}]{lock2017}
{Lock} SJ {Stewart} ST (2017) {The structure of terrestrial bodies: Impact
  heating, corotation limits, and synestias}. Journal of Geophysical Research
  (Planets) 122:950--982

\bibitem[{{Madhusudhan} and {Winn}(2009)}]{madhusudhan2008}
{Madhusudhan} N {Winn} JN (2009) {Empirical Constraints on Trojan Companions
  and Orbital Eccentricities in 25 Transiting Exoplanetary Systems}. \apj
  693:784--793

\bibitem[{{Mamajek} et~al.(2012){Mamajek}, {Quillen}, {Pecaut}, {Moolekamp},
  {Scott}, {Kenworthy}, {Collier Cameron}, and {Parley}}]{mamajek2012}
{Mamajek} EE, {Quillen} AC, {Pecaut} MJ et~al. (2012) {Planetary Construction
  Zones in Occultation: Discovery of an Extrasolar Ring System Transiting a
  Young Sun-like Star and Future Prospects for Detecting Eclipses by
  Circumsecondary and Circumplanetary Disks}. \aj 143:72

\bibitem[{{Marino} et~al.(2017){Marino}, {Wyatt}, {Pani{\'c}}, {Matr{\`a}},
  {Kennedy}, {Bonsor}, {Kral}, {Dent}, {Duchene}, {Wilner}, {Lisse},
  {Lestrade}, and {Matthews}}]{marino2017}
{Marino} S, {Wyatt} MC, {Pani{\'c}} O et~al. (2017) {ALMA observations of the
  {$\eta$} Corvi debris disc: inward scattering of CO-rich exocomets by a chain
  of 3-30 M$_{⊕}$ planets?} \mnras 465:2595--2615

\bibitem[{{McArthur} et~al.(2014){McArthur}, {Benedict}, {Henry}, {Hatzes},
  {Cochran}, {Harrison}, {Johns-Krull}, and {Nelan}}]{mcarthur2014}
{McArthur} BE, {Benedict} GF, {Henry} GW et~al. (2014) {Astrometry, Radial
  Velocity, and Photometry: The HD 128311 System Remixed with Data from HST,
  HET, and APT}. \apj 795:41

\bibitem[{{Miguel} and {Ida}(2016)}]{miguel2016}
{Miguel} Y {Ida} S (2016) {A semi-analytical model for exploring Galilean
  satellites formation from a massive disk}. \icarus 266:1--14

\bibitem[{{Murray} and {Dermott}(2000)}]{murray2000}
{Murray} CD {Dermott} SF (2000) {Solar System Dynamics}

\bibitem[{{Namouni}(2010)}]{namouni2010}
{Namouni} F (2010) {The Fate of Moons of Close-in Giant Exoplanets}. \apjl
  719:L145--L147

\bibitem[{{Nesvorn{\'y}} and {Vokrouhlick{\'y}}(2016)}]{nesvorny2016}
{Nesvorn{\'y}} D {Vokrouhlick{\'y}} D (2016) {Dynamics and Transit Variations
  of Resonant Exoplanets}. \apj 823:72

\bibitem[{{Nesvorn{\'y}} et~al.(2012){Nesvorn{\'y}}, {Kipping}, {Buchhave},
  {Bakos}, {Hartman}, and {Schmitt}}]{nesvorny2012a}
{Nesvorn{\'y}} D, {Kipping} DM, {Buchhave} LA et~al. (2012) {The Detection and
  Characterization of a Nontransiting Planet by Transit Timing Variations}.
  Science 336:1133--

\bibitem[{{Nesvorn{\'y}} et~al.(2013){Nesvorn{\'y}}, {Vokrouhlick{\'y}}, and
  {Morbidelli}}]{nesvorny2013b}
{Nesvorn{\'y}} D, {Vokrouhlick{\'y}} D {Morbidelli} A (2013) {Capture of
  Trojans by Jumping Jupiter}. \apj 768:45

\bibitem[{{Neugebauer} et~al.(2007){Neugebauer}, {Gloeckler}, {Gosling},
  {Rees}, {Skoug}, {Goldstein}, {Armstrong}, {Combi}, {M{\"a}kinen}, {McComas},
  {von Steiger}, {Zurbuchen}, {Smith}, {Geiss}, and
  {Lanzerotti}}]{neugebauer2007}
{Neugebauer} M, {Gloeckler} G, {Gosling} JT et~al. (2007) {Encounter of the
  Ulysses Spacecraft with the Ion Tail of Comet MCNaught}. \apj 667:1262--1266

\bibitem[{{Nielsen} et~al.(2013){Nielsen}, {Liu}, {Wahhaj}, {Biller},
  {Hayward}, {Close}, {Males}, {Skemer}, {Chun}, {Ftaclas}, {Alencar},
  {Artymowicz}, {Boss}, {Clarke}, {de Gouveia Dal Pino}, {Gregorio-Hetem},
  {Hartung}, {Ida}, {Kuchner}, {Lin}, {Reid}, {Shkolnik}, {Tecza}, {Thatte},
  and {Toomey}}]{nielsen2013}
{Nielsen} EL, {Liu} MC, {Wahhaj} Z et~al. (2013) {The Gemini NICI
  Planet-Finding Campaign: The Frequency of Giant Planets around Young B and A
  Stars}. \apj 776:4

\bibitem[{{Ochiai} et~al.(2014){Ochiai}, {Nagasawa}, and {Ida}}]{ochiai2014}
{Ochiai} H, {Nagasawa} M {Ida} S (2014) {Extrasolar Binary Planets. I.
  Formation by Tidal Capture during Planet-Planet Scattering}. \apj 790:92

\bibitem[{{Ogihara} and {Ida}(2012)}]{ogihara2012}
{Ogihara} M {Ida} S (2012) {N-body Simulations of Satellite Formation around
  Giant Planets: Origin of Orbital Configuration of the Galilean Moons}. \apj
  753:60

\bibitem[{{Payne} et~al.(2013){Payne}, {Deck}, {Holman}, and
  {Perets}}]{payne2013}
{Payne} MJ, {Deck} KM, {Holman} MJ {Perets} HB (2013) {Stability of Satellites
  in Closely Packed Planetary Systems}. \apjl 775:L44

\bibitem[{{Placek} et~al.(2015){Placek}, {Knuth}, {Angerhausen}, and
  {Jenkins}}]{placek2015}
{Placek} B, {Knuth} KH, {Angerhausen} D {Jenkins} JM (2015) {Characterization
  of Kepler-91b and the Investigation of a Potential Trojan Companion Using
  EXONEST}. \apj 814:147

\bibitem[{{Poppenhaeger} et~al.(2013){Poppenhaeger}, {Schmitt}, and
  {Wolk}}]{poppenhaeger2013}
{Poppenhaeger} K, {Schmitt} JHMM {Wolk} SJ (2013) {Transit Observations of the
  Hot Jupiter HD 189733b at X-Ray Wavelengths}. \apj 773:62

\bibitem[{{Rabus} et~al.(2009){Rabus}, {Alonso}, {Belmonte}, {Deeg},
  {Gilliland}, {Almenara}, {Brown}, {Charbonneau}, and
  {Mandushev}}]{rabus2009b}
{Rabus} M, {Alonso} R, {Belmonte} JA et~al. (2009) {A cool starspot or a second
  transiting planet in the TrES-1 system?} \aap 494:391--397

\bibitem[{{Rappaport} et~al.(2012){Rappaport}, {Levine}, {Chiang}, {El Mellah},
  {Jenkins}, {Kalomeni}, {Kite}, {Kotson}, {Nelson}, {Rousseau-Nepton}, and
  {Tran}}]{rappaport2012}
{Rappaport} S, {Levine} A, {Chiang} E et~al. (2012) {Possible Disintegrating
  Short-period Super-Mercury Orbiting KIC 12557548}. \apj 752:1

\bibitem[{{Rappaport} et~al.(2014){Rappaport}, {Barclay}, {DeVore}, {Rowe},
  {Sanchis-Ojeda}, and {Still}}]{rappaport2014}
{Rappaport} S, {Barclay} T, {DeVore} J et~al. (2014) {KOI-2700b A Planet
  Candidate with Dusty Effluents on a 22 hr Orbit}. \apj 784:40

\bibitem[{{Rappaport} et~al.(2016){Rappaport}, {Gary}, {Kaye}, {Vanderburg},
  {Croll}, {Benni}, and {Foote}}]{rappaport2016}
{Rappaport} S, {Gary} BL, {Kaye} T et~al. (2016) {Drifting asteroid fragments
  around WD 1145+017}. \mnras 458:3904--3917

\bibitem[{{Rappaport} et~al.(2018){Rappaport}, {Vanderburg}, {Jacobs},
  {LaCourse}, {Jenkins}, {Kraus}, {Rizzuto}, {Latham}, {Bieryla}, {Lazarevic},
  and {Schmitt}}]{rappaport2018}
{Rappaport} S, {Vanderburg} A, {Jacobs} T et~al. (2018) {Likely transiting
  exocomets detected by Kepler}. \mnras 474:1453--1468

\bibitem[{{Rauer} et~al.(2000){Rauer}, {Bockel{\'e}e-Morvan}, {Coustenis},
  {Guillot}, and {Schneider}}]{rauer2000}
{Rauer} H, {Bockel{\'e}e-Morvan} D, {Coustenis} A, {Guillot} T {Schneider} J
  (2000) {Search for an exosphere around 51 Pegasi B with ISO}. \aap
  355:573--580

\bibitem[{{Rauer} et~al.(2014){Rauer}, {Catala}, {Aerts}, {Appourchaux},
  {Benz}, {Brandeker}, {Christensen-Dalsgaard}, {Deleuil}, {Gizon}, {Goupil},
  {G{\"u}del}, {Janot-Pacheco}, {Mas-Hesse}, {Pagano}, {Piotto}, {Pollacco},
  {Santos}, {Smith}, {Su{\'a}rez}, {Szab{\'o}}, {Udry}, {Adibekyan}, {Alibert},
  {Almenara}, {Amaro-Seoane}, {Eiff}, {Asplund}, {Antonello}, {Barnes},
  {Baudin}, {Belkacem}, {Bergemann}, {Bihain}, {Birch}, {Bonfils}, {Boisse},
  {Bonomo}, {Borsa}, {Brand{\~a}o}, {Brocato}, {Brun}, {Burleigh}, {Burston},
  {Cabrera}, {Cassisi}, {Chaplin}, {Charpinet}, {Chiappini}, {Church},
  {Csizmadia}, {Cunha}, {Damasso}, {Davies}, {Deeg}, {D{\'{\i}}az}, {Dreizler},
  {Dreyer}, {Eggenberger}, {Ehrenreich}, {Eigm{\"u}ller}, {Erikson}, {Farmer},
  {Feltzing}, {de Oliveira Fialho}, {Figueira}, {Forveille}, {Fridlund},
  {Garc{\'{\i}}a}, {Giommi}, {Giuffrida}, {Godolt}, {Gomes da Silva},
  {Granzer}, {Grenfell}, {Grotsch-Noels}, {G{\"u}nther}, {Haswell}, {Hatzes},
  {H{\'e}brard}, {Hekker}, {Helled}, {Heng}, {Jenkins}, {Johansen},
  {Khodachenko}, {Kislyakova}, {Kley}, {Kolb}, {Krivova}, {Kupka}, {Lammer},
  {Lanza}, {Lebreton}, {Magrin}, {Marcos-Arenal}, {Marrese}, {Marques},
  {Martins}, {Mathis}, {Mathur}, {Messina}, {Miglio}, {Montalban}, {Montalto},
  {Monteiro}, {Moradi}, {Moravveji}, {Mordasini}, {Morel}, {Mortier},
  {Nascimbeni}, {Nelson}, {Nielsen}, {Noack}, {Norton}, {Ofir}, {Oshagh},
  {Ouazzani}, {P{\'a}pics}, {Parro}, {Petit}, {Plez}, {Poretti}, {Quirrenbach},
  {Ragazzoni}, {Raimondo}, {Rainer}, {Reese}, {Redmer}, {Reffert},
  {Rojas-Ayala}, {Roxburgh}, {Salmon}, {Santerne}, {Schneider}, {Schou},
  {Schuh}, {Schunker}, {Silva-Valio}, {Silvotti}, {Skillen}, {Snellen}, {Sohl},
  {Sousa}, {Sozzetti}, {Stello}, {Strassmeier}, {{\v S}vanda}, {Szab{\'o}},
  {Tkachenko}, {Valencia}, {Van Grootel}, {Vauclair}, {Ventura}, {Wagner},
  {Walton}, {Weingrill}, {Werner}, {Wheatley}, and {Zwintz}}]{rauer2014}
{Rauer} H, {Catala} C, {Aerts} C et~al. (2014) {The PLATO 2.0 mission}.
  Experimental Astronomy 38:249--330

\bibitem[{{Rein}(2015)}]{rein2015}
{Rein} H (2015) {Reanalysis of radial velocity data from the resonant planetary
  system HD128311}. \mnras 448:L58--L61

\bibitem[{{Rein} et~al.(2014){Rein}, {Fujii}, and {Spiegel}}]{rein2014}
{Rein} H, {Fujii} Y {Spiegel} DS (2014) {Some inconvenient truths about
  biosignatures involving two chemical species on Earth-like exoplanets}.
  Proceedings of the National Academy of Science 111:6871--6875

\bibitem[{{Ricker} et~al.(2015){Ricker}, {Winn}, {Vanderspek}, {Latham},
  {Bakos}, {Bean}, {Berta-Thompson}, {Brown}, {Buchhave}, {Butler}, {Butler},
  {Chaplin}, {Charbonneau}, {Christensen-Dalsgaard}, {Clampin}, {Deming},
  {Doty}, {De Lee}, {Dressing}, {Dunham}, {Endl}, {Fressin}, {Ge}, {Henning},
  {Holman}, {Howard}, {Ida}, {Jenkins}, {Jernigan}, {Johnson}, {Kaltenegger},
  {Kawai}, {Kjeldsen}, {Laughlin}, {Levine}, {Lin}, {Lissauer}, {MacQueen},
  {Marcy}, {McCullough}, {Morton}, {Narita}, {Paegert}, {Palle}, {Pepe},
  {Pepper}, {Quirrenbach}, {Rinehart}, {Sasselov}, {Sato}, {Seager},
  {Sozzetti}, {Stassun}, {Sullivan}, {Szentgyorgyi}, {Torres}, {Udry}, and
  {Villasenor}}]{ricker2015}
{Ricker} GR, {Winn} JN, {Vanderspek} R et~al. (2015) {Transiting Exoplanet
  Survey Satellite (TESS)}. Journal of Astronomical Telescopes, Instruments,
  and Systems 1(1):014003

\bibitem[{{Rieder} and {Kenworthy}(2016)}]{rieder2016}
{Rieder} S {Kenworthy} MA (2016) {Constraints on the size and dynamics of the
  J1407b ring system}. \aap 596:A9

\bibitem[{{Sanchis-Ojeda} et~al.(2015){Sanchis-Ojeda}, {Rappaport},
  {Pall{\`e}}, {Delrez}, {DeVore}, {Gandolfi}, {Fukui}, {Ribas}, {Stassun},
  {Albrecht}, {Dai}, {Gaidos}, {Gillon}, {Hirano}, {Holman}, {Howard},
  {Isaacson}, {Jehin}, {Kuzuhara}, {Mann}, {Marcy}, {Miles-P{\'a}ez},
  {Monta{\~n}{\'e}s-Rodr{\'{\i}}guez}, {Murgas}, {Narita}, {Nowak}, {Onitsuka},
  {Paegert}, {Van Eylen}, {Winn}, and {Yu}}]{sanchisojeda2015}
{Sanchis-Ojeda} R, {Rappaport} S, {Pall{\`e}} E et~al. (2015) {The K2-ESPRINT
  Project I: Discovery of the Disintegrating Rocky Planet K2-22b with a
  Cometary Head and Leading Tail}. \apj 812:112

\bibitem[{{Santos} et~al.(2015){Santos}, {Martins}, {Bou{\'e}}, {Correia},
  {Oshagh}, {Figueira}, {Santerne}, {Sousa}, {Melo}, {Montalto}, {Boisse},
  {Ehrenreich}, {Lovis}, {Pepe}, {Udry}, and {Garcia Munoz}}]{santos2015}
{Santos} NC, {Martins} JHC, {Bou{\'e}} G et~al. (2015) {Detecting ring systems
  around exoplanets using high resolution spectroscopy: the case of 51 Pegasi
  b}. \aap 583:A50

\bibitem[{{Sartoretti} and {Schneider}(1999)}]{sartoretti1999}
{Sartoretti} P {Schneider} J (1999) {On the detection of satellites of
  extrasolar planets with the method of transits}. \aaps 134:553--560

\bibitem[{{Sasaki} and {Barnes}(2014)}]{sasaki2014}
{Sasaki} T {Barnes} JW (2014) {Longevity of moons around habitable planets}.
  International Journal of Astrobiology 13:324--336

\bibitem[{{Sasaki} et~al.(2012){Sasaki}, {Barnes}, and {O'Brien}}]{sasaki2012}
{Sasaki} T, {Barnes} JW {O'Brien} DP (2012) {Outcomes and Duration of Tidal
  Evolution in a Star-Planet-Moon System}. \apj 754:51

\bibitem[{{Scharf}(2006)}]{scharf2006}
{Scharf} CA (2006) {The Potential for Tidally Heated Icy and Temperate Moons
  around Exoplanets}. \apj 648:1196--1205

\bibitem[{{Schneider} et~al.(2010){Schneider}, {L{\'e}ger}, {Fridlund},
  {White}, {Eiroa}, {Henning}, {Herbst}, {Lammer}, {Liseau}, {Paresce},
  {Penny}, {Quirrenbach}, {R{\"o}ttgering}, {Selsis}, {Beichman}, {Danchi},
  {Kaltenegger}, {Lunine}, {Stam}, and {Tinetti}}]{schneider2010}
{Schneider} J, {L{\'e}ger} A, {Fridlund} M et~al. (2010) {The Far Future of
  Exoplanet Direct Characterization}. Astrobiology 10:121--126

\bibitem[{{Schneider} et~al.(2015){Schneider}, {Lainey}, and
  {Cabrera}}]{schneider2015}
{Schneider} J, {Lainey} V {Cabrera} J (2015) {A next step in exoplanetology:
  exo-moons}. International Journal of Astrobiology 14:191--199

\bibitem[{{Simon} et~al.(2015){Simon}, {Szab{\'o}}, {Kiss}, {Fortier}, and
  {Benz}}]{simon2015}
{Simon} AE, {Szab{\'o}} GM, {Kiss} LL, {Fortier} A {Benz} W (2015) {CHEOPS
  Performance for Exomoons: The Detectability of Exomoons by Using Optimal
  Decision Algorithm}. \pasp 127:1084

\bibitem[{{Sinukoff} et~al.(2013){Sinukoff}, {Fulton}, {Scuderi}, and
  {Gaidos}}]{sinukoff2013}
{Sinukoff} E, {Fulton} B, {Scuderi} L {Gaidos} E (2013) {Below One Earth: The
  Detection, Formation, and Properties of Subterrestrial Worlds}. \ssr
  180:71--99

\bibitem[{{Skowron} et~al.(2014){Skowron}, {Udalski}, {Szyma{\'n}ski},
  {Kubiak}, {Pietrzy{\'n}ski}, {Soszy{\'n}ski}, {Poleski}, {Ulaczyk},
  {Pietrukowicz}, {Koz{\l}owski}, {Wyrzykowski}, and {Gould}}]{skowron2014}
{Skowron} J, {Udalski} A, {Szyma{\'n}ski} MK et~al. (2014) {New Method to
  Measure Proper Motions of Microlensed Sources: Application to Candidate
  Free-floating-planet Event MOA-2011-BLG-262}. \apj 785:156

\bibitem[{{Spalding} et~al.(2016){Spalding}, {Batygin}, and
  {Adams}}]{spalding2016}
{Spalding} C, {Batygin} K {Adams} FC (2016) {Resonant Removal of Exomoons
  during Planetary Migration}. \apj 817:18

\bibitem[{{Spergel} et~al.(2015){Spergel}, {Gehrels}, {Baltay}, {Bennett},
  {Breckinridge}, {Donahue}, {Dressler}, {Gaudi}, {Greene}, {Guyon}, {Hirata},
  {Kalirai}, {Kasdin}, {Macintosh}, {Moos}, {Perlmutter}, {Postman},
  {Rauscher}, {Rhodes}, {Wang}, {Weinberg}, {Benford}, {Hudson}, {Jeong},
  {Mellier}, {Traub}, {Yamada}, {Capak}, {Colbert}, {Masters}, {Penny},
  {Savransky}, {Stern}, {Zimmerman}, {Barry}, {Bartusek}, {Carpenter}, {Cheng},
  {Content}, {Dekens}, {Demers}, {Grady}, {Jackson}, {Kuan}, {Kruk}, {Melton},
  {Nemati}, {Parvin}, {Poberezhskiy}, {Peddie}, {Ruffa}, {Wallace}, {Whipple},
  {Wollack}, and {Zhao}}]{spergel2015wfirst}
{Spergel} D, {Gehrels} N, {Baltay} C et~al. (2015) {Wide-Field InfrarRed Survey
  Telescope-Astrophysics Focused Telescope Assets WFIRST-AFTA 2015 Report}.
  ArXiv e-prints

\bibitem[{{Szab{\'o}} et~al.(2006){Szab{\'o}}, {Szatm{\'a}ry}, {Div{\'e}ki},
  and {Simon}}]{szabo2006}
{Szab{\'o}} GM, {Szatm{\'a}ry} K, {Div{\'e}ki} Z {Simon} A (2006) {Possibility
  of a photometric detection of ''exomoons''}. \aap 450:395--398

\bibitem[{{Szab{\'o}} et~al.(2013){Szab{\'o}}, {Szab{\'o}}, {D{\'a}lya},
  {Simon}, {Hodos{\'a}n}, and {Kiss}}]{szabo2013}
{Szab{\'o}} R, {Szab{\'o}} GM, {D{\'a}lya} G et~al. (2013) {Multiple planets or
  exomoons in Kepler hot Jupiter systems with transit timing variations?} \aap
  553:A17

\bibitem[{{Teachey} et~al.(2018){Teachey}, {Kipping}, and
  {Schmitt}}]{teachey2018}
{Teachey} A, {Kipping} DM {Schmitt} AR (2018) {HEK. VI. On the Dearth of
  Galilean Analogs in Kepler, and the Exomoon Candidate Kepler-1625b I}. \aj
  155:36

\bibitem[{{Turner} et~al.(2016){Turner}, {Pearson}, {Biddle}, {Smart},
  {Zellem}, {Teske}, {Hardegree-Ullman}, {Griffith}, {Leiter}, {Cates},
  {Nieberding}, {Smith}, {Thompson}, {Hofmann}, {Berube}, {Nguyen}, {Small},
  {Guvenen}, {Richardson}, {McGraw}, {Raphael}, {Crawford}, {Robertson},
  {Tombleson}, {Carleton}, {Towner}, {Walker-LaFollette}, {Hume}, {Watson},
  {Jones}, {Lichtenberger}, {Hoglund}, {Cook}, {Crossen}, {Jorgensen},
  {Romine}, {Thompson}, {Villegas}, {Wilson}, {Sanford}, {Taylor}, and
  {Henz}}]{turner2016}
{Turner} JD, {Pearson} KA, {Biddle} LI et~al. (2016) {Ground-based near-UV
  observations of 15 transiting exoplanets: constraints on their atmospheres
  and no evidence for asymmetrical transits}. \mnras 459:789--819

\bibitem[{{Uehara} et~al.(2016){Uehara}, {Kawahara}, {Masuda}, {Yamada}, and
  {Aizawa}}]{uehara2016}
{Uehara} S, {Kawahara} H, {Masuda} K, {Yamada} S {Aizawa} M (2016) {Transiting
  Planet Candidates Beyond the Snow Line Detected by Visual Inspection of 7557
  Kepler Objects of Interest}. \apj 822:2

\bibitem[{{van Lieshout} et~al.(2014){van Lieshout}, {Min}, and
  {Dominik}}]{vanlieshout2014}
{van Lieshout} R, {Min} M {Dominik} C (2014) {Dusty tails of evaporating
  exoplanets. I. Constraints on the dust composition}. \aap 572:A76

\bibitem[{{van Lieshout} et~al.(2016){van Lieshout}, {Min}, {Dominik}, {Brogi},
  {de Graaff}, {Hekker}, {Kama}, {Keller}, {Ridden-Harper}, and {van
  Werkhoven}}]{vanlieshout2016}
{van Lieshout} R, {Min} M, {Dominik} C et~al. (2016) {Dusty tails of
  evaporating exoplanets. II. Physical modelling of the KIC 12557548b light
  curve}. \aap 596:A32

\bibitem[{{Vanderburg} et~al.(2015){Vanderburg}, {Johnson}, {Rappaport},
  {Bieryla}, {Irwin}, {Lewis}, {Kipping}, {Brown}, {Dufour}, {Ciardi}, {Angus},
  {Schaefer}, {Latham}, {Charbonneau}, {Beichman}, {Eastman}, {McCrady},
  {Wittenmyer}, and {Wright}}]{vanderburg2015}
{Vanderburg} A, {Johnson} JA, {Rappaport} S et~al. (2015) {A disintegrating
  minor planet transiting a white dwarf}. \nat 526:546--549

\bibitem[{{Veras} et~al.(2016){Veras}, {Marsh}, and {G{\"a}nsicke}}]{veras2016}
{Veras} D, {Marsh} TR {G{\"a}nsicke} BT (2016) {Dynamical mass and multiplicity
  constraints on co-orbital bodies around stars}. \mnras 461:1413--1420

\bibitem[{{Vidal-Madjar} et~al.(2003){Vidal-Madjar}, {Lecavelier des Etangs},
  {D{\'e}sert}, {Ballester}, {Ferlet}, {H{\'e}brard}, and
  {Mayor}}]{vidalmadjar2003}
{Vidal-Madjar} A, {Lecavelier des Etangs} A, {D{\'e}sert} JM et~al. (2003) {An
  extended upper atmosphere around the extrasolar planet HD209458b}. \nat
  422:143--146

\bibitem[{{Vokrouhlick{\'y}} and {Nesvorn{\'y}}(2014)}]{vokrouhlick2014}
{Vokrouhlick{\'y}} D {Nesvorn{\'y}} D (2014) {Transit Timing Variations for
  Planets Co-orbiting in the Horseshoe Regime}. \apj 791:6

\bibitem[{Waite et~al.(2017)Waite, Glein, Perryman, Teolis, Magee, Miller,
  Grimes, Perry, Miller, Bouquet, Lunine, Brockwell, and Bolton}]{waite2017}
Waite JH, Glein CR, Perryman RS et~al. (2017) Cassini finds molecular hydrogen
  in the enceladus plume: Evidence for hydrothermal processes. Science
  356(6334):155--159,
  \urlprefix\url{http://science.sciencemag.org/content/356/6334/155}

\bibitem[{{Wang} et~al.(2016){Wang}, {Graham}, {Pueyo}, {Kalas},
  {Millar-Blanchaer}, {Ruffio}, {De Rosa}, {Ammons}, {Arriaga}, {Bailey},
  {Barman}, {Bulger}, {Burrows}, {Cardwell}, {Chen}, {Chilcote}, {Cotten},
  {Fitzgerald}, {Follette}, {Doyon}, {Duch{\^e}ne}, {Greenbaum}, {Hibon},
  {Hung}, {Ingraham}, {Konopacky}, {Larkin}, {Macintosh}, {Maire}, {Marchis},
  {Marley}, {Marois}, {Metchev}, {Nielsen}, {Oppenheimer}, {Palmer}, {Patel},
  {Patience}, {Perrin}, {Poyneer}, {Rajan}, {Rameau}, {Rantakyr{\"o}},
  {Savransky}, {Sivaramakrishnan}, {Song}, {Soummer}, {Thomas}, {Vasisht},
  {Vega}, {Wallace}, {Ward-Duong}, {Wiktorowicz}, and {Wolff}}]{wang2016}
{Wang} JJ, {Graham} JR, {Pueyo} L et~al. (2016) {The Orbit and Transit
  Prospects for {$\beta$} Pictoris b Constrained with One Milliarcsecond
  Astrometry}. \aj 152:97

\bibitem[{{Weidner} and {Horne}(2010)}]{weidner2010}
{Weidner} C {Horne} K (2010) {Limits on the orbits and masses of moons around
  currently-known transiting exoplanets}. \aap 521:A76

\bibitem[{{Wertz} et~al.(2017){Wertz}, {Absil}, {G{\'o}mez Gonz{\'a}lez},
  {Milli}, {Girard}, {Mawet}, and {Pueyo}}]{wertz2017}
{Wertz} O, {Absil} O, {G{\'o}mez Gonz{\'a}lez} CA et~al. (2017) {VLT/SPHERE
  robust astrometry of the HR8799 planets at milliarcsecond-level accuracy.
  Orbital architecture analysis with PyAstrOFit}. \aap 598:A83

\bibitem[{{Wright} and {Gaudi}(2013)}]{wright2012}
{Wright} JT {Gaudi} BS (2013) Exoplanet detection methods. In: {Oswalt} T,
  {French} LM {Kalas} P (eds) Planets, Stars and Stellar Systems, vol~3, pp
  489--540

\bibitem[{{Wright} and {Sigurdsson}(2016)}]{wright2016}
{Wright} JT {Sigurdsson} S (2016) {Families of Plausible Solutions to the
  Puzzle of Boyajians Star}. \apjl 829:L3

\bibitem[{{Yang} et~al.(2016){Yang}, {Xie}, {Zhou}, {Liu}, and
  {Zhang}}]{yang2016}
{Yang} M, {Xie} JW, {Zhou} JL, {Liu} HG {Zhang} H (2016) {Global Instability of
  the Exo-moon System Triggered by Photo-evaporation}. \apj 833:7

\bibitem[{{Zakhozhay} et~al.(2017){Zakhozhay}, {Zapatero Osorio}, {B{\'e}jar},
  and {Boehler}}]{zakhozhay2017}
{Zakhozhay} OV, {Zapatero Osorio} MR, {B{\'e}jar} VJS {Boehler} Y (2017)
  {Spectral energy distribution simulations of a possible ring structure around
  the young, red brown dwarf G 196-3 B}. \mnras 464:1108--1118

\bibitem[{{Zanazzi} and {Lai}(2017)}]{zanazzi2017}
{Zanazzi} JJ {Lai} D (2017) {Extended transiting discs and rings around planets
  and brown dwarfs: theoretical constraints}. \mnras 464:3945--3954

\bibitem[{{Zhou} et~al.(2016){Zhou}, {Kedziora-Chudczer}, {Bailey}, {Marshall},
  {Bayliss}, {Stockdale}, {Nelson}, {Tan}, {Rodriguez}, {Tinney}, {Dragomir},
  {Colon}, {Shporer}, {Bento}, {Sefako}, {Horne}, and {Cochran}}]{zhou2016}
{Zhou} G, {Kedziora-Chudczer} L, {Bailey} J et~al. (2016) {Simultaneous
  infrared and optical observations of the transiting debris cloud around WD
  1145+017}. \mnras 463:4422--4432

\end{thebibliography}

\end{document}